\documentclass[Crown,sagev,times]{sagej}

\usepackage{moreverb,url}

\usepackage[colorlinks,bookmarksopen,bookmarksnumbered,citecolor=red,urlcolor=red]{hyperref}

\newcommand\BibTeX{{\rmfamily B\kern-.05em \textsc{i\kern-.025em b}\kern-.08em
T\kern-.1667em\lower.7ex\hbox{E}\kern-.125emX}}

\def\volumeyear{2022}
\usepackage{float}
\usepackage{epsfig}
\usepackage{amssymb}
\usepackage{amsmath}
\usepackage{latexsym}
\usepackage{color}
\font\tenbf=cmbx9

\font\tenrm=cmr9 

\font\tenit=cmti9

\font\sc=cmr12

\newcommand{\lle}{\mbox{$\langle$}}
\newcommand{\rle}{\mbox{$\rangle$}}

\newcommand{\bfsi}{\mbox{\boldmath$\sigma$}}

\newcommand{\bfep}{\mbox{\boldmath$\varepsilon$}}

\newcommand{\bfze}{\mbox{\boldmath$\zeta$}}

\newcommand{\bfchi}{\mbox{\boldmath$\chi$}}

\newcommand{\bfcK}{\mbox{\boldmath$\cal K$}}
\newcommand{\bfcL}{\mbox{\boldmath$\cal L$}}
\newcommand{\bfcG}{\mbox{\boldmath$\cal G$}}

\newcommand{\bfcU}{\mbox{\boldmath$\cal U$}}

\newcommand{\bfb}{\mbox{\boldmath$\bf b$}}

\newcommand{\bfe}{\mbox{\boldmath$\bf e$}}

\newcommand{\bfm}{\mbox{\boldmath$\bf m$}}
\newcommand{\bff}{\mbox{\boldmath$\bf f$}}
\newcommand{\bfh}{\mbox{\boldmath$\bf h$}}
\newcommand{\bfg}{\mbox{\boldmath$\bf g$}}
\newcommand{\bfk}{\mbox{\boldmath$\bf k$}}

\newcommand{\bfn}{\mbox{\boldmath$\bf n$}}
\newcommand{\bfp}{\mbox{\boldmath$\bf p$}}

\newcommand{\bft}{\mbox{\boldmath$\bf t$}}
\newcommand{\bfs}{\mbox{\boldmath$\bf s$}}
\newcommand{\bfr}{\mbox{\boldmath$\bf r$}}
\newcommand{\bfu}{\mbox{\boldmath$\bf u$}}

\newcommand{\bfw}{\mbox{\boldmath$\bf w$}}
\newcommand{\bfx}{\mbox{\boldmath$\bf x$}}
\newcommand{\bfy}{\mbox{\boldmath$\bf y$}}
\newcommand{\bfz}{\mbox{\boldmath$\bf z$}}
\newcommand{\bfA}{\mbox{\boldmath$\bf A$}}

\newcommand{\bfG}{\mbox{\boldmath$\bf G$}}

\newcommand{\bfI}{\mbox{\boldmath$\bf I$}}

\newcommand{\bfL}{\mbox{\boldmath$\bf L$}}
\newcommand{\bfN}{\mbox{\boldmath$\bf N$}}

\newcommand{\bfX}{\mbox{\boldmath$\bf X$}}
\newcommand{\bfY}{\mbox{\boldmath$\bf Y$}}

\newcommand{\bfU}{\mbox{\boldmath$\bf U$}}
\newcommand{\bfF}{\mbox{\boldmath$\bf F$}}
\newcommand{\bfR}{\mbox{\boldmath$\bf R$}}
\newcommand{\bfK}{\mbox{\boldmath$\bf K$}}
\newcommand{\bfM}{\mbox{\boldmath$\bf M$}}

\newcommand{\bfdelta}{\mbox{\boldmath$\delta$}}

\newcommand{\bfxi}{\mbox{\boldmath$\xi$}}

\newcommand{\cF}{\mbox{$\cal F$}}

\newcommand{\bfbd}{\mbox{$\mathfrak{d}$}}
\newcommand{\bfbs}{\mbox{$\mathfrak{s}$}}

\newcommand{\bfLa}{\mbox{\boldmath$\Lambda$}}
\newcommand{\bfcD}{\mbox{\boldmath$\cal D$}}
\newcommand{\bfcI}{\mbox{\boldmath$\cal I$}}

\newcommand{\bfcF}{\mbox{\boldmath$\cal F$}}

\newcommand{\bfcA}{\mbox{\boldmath$\cal A$}}
\newcommand{\bfcB}{\mbox{\boldmath$\cal B$}}

\newcommand{\cV}{\mbox{$\cal V$}}

\newcommand{\bftau}{\mbox{\boldmath$\tau$}}

\newcommand{\bfthe}{\mbox{\boldmath$\vartheta$}}

\newcommand{\BB}{\begin{equation}}
\newcommand{\EE}{\end{equation}}

\newcommand{\BBEQ}{\begin{eqnarray}}
\newcommand{\EEEQ}{\end{eqnarray}}

\begin{document}

\runninghead{Buryachenko}

\title{New RVE concept and FFT methods in micromechanics of composites
subjected to body force with compact support } %% Article title

\author{Valeriy A. Buryachenko\affilnum{1}}

\affiliation{\affilnum{1}Micromechanics and Composites LLC, Cincinnati, OH 45202, USA} 

\corrauth{Valeriy A. Buryachenko, Cincinnati, OH 45202, USA.}

\email{buryach@yahoo.com}

%% Abstract
\begin{abstract}

We consider static linear elastic composite materials (CMs) with periodic structure. The core of the proposed methodology is the generation of a novel dataset using specially designed body force fields with compact support (BFCS), enabling a new RVE concept that reduces the infinite periodic medium to a finite domain without boundary artifacts. This functionally reduced RVE is used for translated averaging of direct numerical simulations (DNS) results, efficiently computed via a newly developed FFT-based solver for BFCS loading.
The resulting dataset captures localized field responses and is used to train machine learning (ML) and neural networks (NN) models to learn effective nonlocal surrogate operators. These operators accurately predict macroscopic responses while reflecting microstructural features and nonlocal interactions. By accounting for field localization  while simultaneously eliminating influences from finite sample size and boundary effects, it provides a physically grounded and  data-driven framework for constructing accurate surrogate models for the homogenization of complex materials.
\end{abstract}


\maketitle
\section{Introduction }

Periodic composites, due to their inherent microstructural regularity, are well-suited for multiscale homogenization frameworks  \citep{{Fish`2014},  {Ghosh`2011},  {Zohdi`W`2008}}. In asymptotic homogenization, originally developed by Babuška and further formalized in  \citep{{Bakhvalov`P`1984},  {Fish`2014}}, the response of a periodic medium is approximated by separating scales under the assumption that the unit cell is much smaller than the overall material size that leads to homogenized coefficients via unit-cell problems.
Conversely, computational homogenization methods resolve the microstructural field equations numerically, enabling the modeling of nonlocal \citep{Buryachenko`2024} and inelastic behavior  \citep{{Kouznetsova`et`2001},  {Matous`et`2017},  {Terada`K`2001}}. A prominent framework is the FE$^2$ scheme, where the macroscopic finite element problem is coupled to microscale RVEs embedded at each Gauss integration point  \citep{{Geers`et`2010},  {Kanout`et`2009},  {Raju`et`2021}}. Each RVE solves a boundary value problem consistent with macroscopic deformation modes, typically under periodic boundary conditions.
While FE$^2$ provides a rigorous scale-bridging strategy, its computational cost is significant due to the nested finite element discretizations, often requiring high-performance computing or model order reduction techniques for tractability in large-scale simulations.

FFT-based numerical homogenization techniques, first introduced by Moulinec and Suquet  \citep{{Moulinec`S`1994},  {Moulinec`S`1998}}, offer a computationally efficient alternative to finite element methods (FEM), reducing complexity from $O(N^2)$ to $O(N {\rm log} N)$. via spectral solvers that leverage the discrete Fourier transform (DFT). The original scheme relies on the solution of the periodic Lippmann–Schwinger (L–S) integral equation, reformulated in Fourier space where convolution with the Green’s operator becomes an element-wise multiplication, enabling efficient iterative solvers.
FFT methods are generally grouped into three technical categories:
1). Lippmann–Schwinger-type solvers: These include polarization-based iterative schemes and Krylov-subspace methods (e.g., conjugate gradient, BiCG) for solving the linearized form of the L–S equation  \citep{{Michel`et`2001},  {Brisard`D`2010},  {Zeman`et`2010},  {Wicht`et`2021},  {Wang`et`2024}}. Recent developments have incorporated machine learning—e.g., FFT-NN hybrids for multiscale prediction in complex woven composites  \citep{LiM`et`2024}.
2). Fourier–Galerkin methods: Derived from a variational (weak-form) framework, these approaches utilize trigonometric polynomial basis functions with consistent projection operators to enhance convergence and accuracy, particularly for materials with discontinuities or high contrast  \citep{{deGeus`et`2017},  {Zeman`et`2017}}.
3). Displacement-based FFT solvers: These formulations treat the displacement fluctuation field as the primary unknown and solve equilibrium equations directly in Fourier space, avoiding the need for a reference medium and facilitating the use of preconditioned Krylov solvers  \citep{Lucarini`S`2019}.
Advanced FFT-based solvers are now used in highly nonlinear regimes (e.g., elastoplasticity  \citep{Schneider`2021}) and in the context of crystal plasticity  \citep{Segurado`et`2018}. Many implementations are open-source and exploit domain decomposition and parallel FFT libraries to scale across distributed memory architectures  \citep{Lucarini`et`2022}, achieving optimal $O(N {\rm log} N)$ performance in large-scale simulations.

The representative volume element (RVE) is fundamental for predicting the effective behavior of heterogeneous materials. It must be large enough to capture the essential microstructural features while ensuring that macroscopic responses are independent of boundary conditions and representative of the bulk material. Hill's classical definition  \citep{Hill`1963} requires macroscopically uniform boundary conditions and effective properties described by homogenized moduli.
RVE size selection involves achieving scale separation, where the microstructural length scale $a$ satisfies $a\ll \Lambda\ll L$, with 
$\Lambda$ being the applied field scale and $L$ the macroscale domain length. 
The minimal domain size where effective properties stabilize is taken as the RVE. Related to this is the concept of the statistically equivalent RVE (SERVE), which uses micro-computed tomography (micro-CT) imaging and statistical analysis to construct realistic computational domains. For detailed methodologies, see  \citep{{Bargmann`et`2018},  {Kanit`et`2003},  {Matous`et`2017},  {Moumen`et`2021},  {Ostoja`et`2016}}.

When scale separation is violated, statistical homogeneity breaks down, leading to nonlocal coupling between stress and strain fields. This interaction is governed by a tensorial kernel, requiring the use of effective nonlocal operators—either integral or higher-order differential forms—instead of classical effective moduli  \citep{Hill`1963}. These operators capture the influence of distant points in the material. Nonlocal models fall into two categories: strongly nonlocal (such as strain-based or displacement-based methods like peridynamics) and weakly nonlocal (e.g., strain- or stress-gradient theories). {\color{black} This shift necessitates redefining the representative volume element (RVE) to correspond with the chosen nonlocal operator, applicable to both random \citep{{Drugan`2000},  {Drugan`2003},  {Drugan`W`1996}} and periodic \citep{{Ameen`et`2018},  {Kouznetsova`et`2004a},  {Kouznetsova`et`2004b},  {Smyshlyaev`C`2000}} composites. 
These generalized RVEs are defined as the minimal domain size at which a predefined effective (nonlocal) operator—unlike the classical effective moduli of \citep{Hill`1963}—stabilizes.}
The RVE becomes even more critical when accounting for combined nonlocal effects from boundary conditions, intrinsic material behavior, and phase interactions.

The evolution of effective nonlocal operator theory has been greatly accelerated by the integration of machine learning (ML) and neural networks (NN), introducing new levels of flexibility and modeling power. Initial efforts by Silling  \citep{Silling`2020} and You 
{\it et al}.  \citep{{You`et`2020},  {You`et`2024}} demonstrated how DNS data could be used to construct surrogate integral operators for complex materials. More recently, nonlocal neural operators have emerged as tools for learning mappings between function spaces  \citep{{Li`et`2003},  {Lanthaler`et`2024}}. 
Numerous neural operator architectures have been developed, including DeepONet, PCA-Net, Graph Neural Operators, FNO, and LNO, each tailored to different aspects of operator learning. Comparative reviews can be found in  \citep{{Gosmani`et`2022}, 
 {Hu`et`2024},  {Kumara`Y`2023}, {Lanthaler`et`2024}}. For nonlocal mechanics, the Peridynamic Neural Operator (PNO)  \citep{Jafarzadeh`et`2024} and its heterogeneous extension HeteroPNO \citep{Jafarzadeh`et`2024b} provide physics-aware modeling of peridynamic interactions. 
Physics-Informed Neural Networks (PINNs) further enhance NN models by embedding governing equations directly as soft constraints  \citep{{Raissi`et`2019}, 
 {Karniadakis`et`2021},  {Hu`et`2024}}, ensuring consistency with physical laws. When neural operators are combined with PINNs  \citep{{Faroughi`et`2024},  {Gosmani`et`2022},  {Wang`Y`2024}}, the resulting frameworks can accurately model complex nonlinear, heterogeneous, and nonlocal material behavior with strong generalization capabilities.

While ML and NN methods have advanced material modeling significantly, they frequently miss key micromechanical considerations—such as scale separation, boundary effects, and the RVE--which are essential for accurate predictions in both linear and nonlinear regimes. To bridge this gap, the proposed approach constructs new types of compressed datasets for complex microstructures (random or periodic), based on an innovative RVE framework. 
This novel RVE concept is independent of the constitutive behavior of individual phases and the analytical form of surrogate operators. Instead, it leverages field concentration factors within each phase to characterize the microstructure effectively. The resulting datasets, enriched by this micromechanically informed RVE, should be compatible with any ML or NN architecture for predicting nonlocal surrogate operators. This innovative RVE concept ensures the accuracy of predictions by removing potential issues related to size scale, boundary layers, and edge effects.

The proposed approach is composed of several key components, some of which are fundamentally novel and mark a significant departure from traditional techniques. At the core of the methodology is the generation of a new type of dataset, derived through the application of specifically designed body force fields with compact support. These body forces are spatially localized, allowing for precise control of excitation within the material sample. This feature enables the formulation of a new RVE concept, in which the infinite periodic microstructure is effectively reduced to a finite computational domain, without sacrificing the fidelity of microstructural response or introducing boundary-related artifacts.
This RVE reduction is not merely geometric, but functional: the dataset is built through translated averaging of direct numerical simulations (DNS) performed within this finite domain. To enable efficient simulation, a newly developed FFT-based solver tailored for CMs is employed. This FFT solver is adapted to handle the response of materials subjected to body forces with compact support, leading to fast, scalable, and accurate computation of the local fields.
Once this enhanced dataset is constructed—containing information that is rich in both spatial resolution and microstructural mechanics—it serves as a training ground for ML and NN models. These models are trained to learn the effective surrogate nonlocal operators that can predict the material response under arbitrary macroscopic loading conditions. Importantly, the surrogate operator constructed via this framework reflects both the nonlocal interactions inherent in the material and the fine-scale features encoded in the dataset. Therefore, the proposed methodology effectively integrates rigorous micromechanical principles with advanced data-driven modeling techniques. By accounting for field localization while simultaneously eliminating influences from finite sample size and boundary effects, it provides a physically grounded and broadly applicable framework for constructing accurate surrogate models for the homogenization of complex materials.

The structure of the paper is organized as follows. Section 2 introduces both the classical and modified forms of the Lippmann–Schwinger equations. Section 3 presents the traditional and newly proposed RVE concepts, along with the methodology for selecting the corresponding datasets. A brief overview of the FFT framework, adapted to support later developments, is provided in Section 4. In Section 5, new FFT-based methods are introduced for analyzing composite materials (CMs) under body force fields with compact support (BFCS). Finally, Section 6 details how the newly generated datasets are integrated into existing machine learning (ML) and neural network (NN) approaches for constructing surrogate models.

\section{Modified Lippmann–Schwinger equation}
\setcounter{equation}{0}
\renewcommand{\theequation}{2.\arabic{equation}}

We consider a linear elastic body occupying an open simply connected bounded domain $w\subset R^d$
with a smooth boundary $\Gamma_0$ and with an indicator function $W$ and space dimensionality
$d$ ($d=2$ and $d=3$ for 2-$D$ and 3-$D$ problems, respectively).
The domain $w$ {\color{black} with the boundary $\Gamma^0$} contains a homogeneous matrix $v^{(0)}$ and a periodic
{\color{black} field} $X=(v_i)$ of heterogeneity $v_i$ with the centers $\bfx_i$, indicator functions $V_i$ and bounded by the closed
smooth surfaces $\Gamma_i$ $(i=1,2,\ldots)$.
It is presumed that the heterogeneities can be grouped into phases $v^{(q)} \quad (q=1,2,\ldots,N)$ with identical mechanical and geometrical properties.
We first consider the local basic equations of thermoelasticity of composites
\BBEQ
\label{2.1}
{\color{black}\nabla\cdot}\bfsi(\bfx)&=&-\bfb(\bfx), \\ %(2.1)
\label{2.2}
\bfsi(\bfx)&=&\bfL(\bfx)\bfep(\bfx), \ \ {\rm or}\ \ \
\bfep(\bfx)=\bfM(\bfx)\bfsi(\bfx), \\ %(2.2)
\label{2.3}
\bfep(\bfx)&=&\nabla^s\bfu, \ \
%+(\nabla {\bf u})^{\top}]/2,\ \
\nabla\times\bfep(\bfx)\times\nabla={\bf 0}, %(2.3)
\EEEQ
where $\otimes$ and
and $\times$ are the tensor and vector products, respectively, 
$\nabla^s$ is the symmetric gradient operator, 
$\nabla^s\bfu:=[\nabla {\otimes}{\bf u}+(\nabla{\otimes}{\bf u})^{\top}]/2$, and $(.)^{\top}$ denotes matrix transposition;
$\bfb$ is the body force.
${\bf L(x)}$ and ${\bf M(x) \equiv L(x)}^{-1}$ are the known phase
stiffness and compliance tensors.
In particular, for isotropic
constituents, the local stiffness tensor $\bfL(\bfx)$ is given in
terms of the local bulk modulus $k(\bfx)$ and the local shear
modulus $\mu(\bfx)$ and:
\BB
%\label{2.4}
\bfL(\bfx)=(dk,2\mu)\equiv dk(\bfx)\bfN_1+2\mu(\bfx)\bfN_2, \nonumber
%(2.4)
\EE
${\bf N}_1=\bfdelta\otimes\bfdelta/d, \ {\bf N}_2={\bf I-N}_1$ $(d=2\ {\rm or}\ 3$) whereas
$\bfdelta$ and $\bfI$ are the unit second-order and fourth-order tensors.
For all material tensors $\bfg$ (e.g., $\bfL, \bfM)$ the notation $\bfg_1(\bfx)\equiv \bfg(\bfx)-\bfg^{(0)}=\bfg^{(m)}_1(\bfx)$ $(\bfx\in v^{(m)},\ m=0,1,\ldots, N$) is used.
The upper index $^{(m)}$ indicates the
components, and the lower index $i$ shows the individual
heterogeneity; $v^{(0)}=w\backslash v$, $ v\equiv \cup v^{(k)}\equiv\cup v_i,
\ V(\bfx)=\sum V^{(k)}=\sum V_i(\bfx)$, and $V^{(k)}(\bfx)$ and $V_i(\bfx)$ are the
indicator functions of $v^{(k)}$ and $v_i$, respectively, equals 1 at
$\bfx\in v^{(k)}$ and 0 otherwise, $(m=0,k;\ k=1,2,\ldots,N;
\quad i=1,2,\ldots)$.

Substituting Eqs. (\ref{2.2}) and (\ref{2.3}$_1$) into Eq. (\ref{2.1}) leads to the equilibrium equation (\ref{2.1}) being expressed in the form:
\BB
\label{2.4}
{\bfcL}(\bfu)(\bfx)+\bfb(\bfx)={\bf 0},\ \ \ {\bfcL}(\bfu)(\bfx):=\nabla[\bfL\nabla\bfu(\bfx)],
\EE
with ${\bfcL}(\bfu)(\bfx)$ representing a second-order elliptic differential operator.

The body force density $\bfb(\bfx)$ is assumed to have compact support, be self-equilibrated, and vanish outside a specified loading region $\bfcB^b:=b({\bf 0}, B^b)$:
\BB
\label{2.5}
\int \bfb(\bfx)d\bfx={\bf 0}, \ \ \ \bfb(\bfy)\equiv {\bf 0}\ \ {\rm for}\ \ \bfy\not\in b({\bf 0}, B^b):=\{\bfy| |\bfy|\leq B^b\},
\EE
where $b({\bf 0}, B^b)$ denotes a ball of radius $B^b$ centered at the origin $\bfx = \mathbf{0}$.

A linear-elastic reference material is introduced, characterized as homogeneous and isotropic with a stiffness tensor $\bfL^{(0)}$. Consider the governing equation for an infinite homogeneous medium occupying $\mathbb{R}^d$ ($d = 1, 2, 3$), subject to the body force density $\bfb(\bfx)$ defined in Eq. (\ref{2.5}):
\BB
\label{2.6}
{\bfcL}^{(0)}(\bfu^{b(0)})(\bfx)+\bfb(\bfx)={\bf 0},
\EE
where ${\bfcL}^{(0)}$ denotes the elliptic operator associated with the homogeneous stiffness tensor $\bfL^{(0)}$.

The body force density with compact support (BFCS) $\bfb(\bfx)$ (\ref{2.5}) is assumed to be self-equilibrated and periodic with respect to a body force unit cell (BFUC) $\Omega^b_{00}\supset \bfcB^b$. 
For notational simplicity in describing periodic BFCS, we restrict our attention to the two-dimensional case, where the entire domain is represented as a union of square unit cells, $w=\cup \Omega_{ij}^b$ ($i,j=0,\pm 1, \pm 2,\ldots$), with the grid of centers 
$\bfLa^b=\{\bfx^{b\Lambda}\}$. 
Let $\Omega_{00}^b$ denote a representative body force unit cell (BFUC), bounded by corner points $\bfx^{bc}_{kl}$ ($k,l = \pm1$), and with boundary $\Gamma^{b0} = \cup \Gamma^{b0}_{ij}$. Each segment $\Gamma^{b0}_{ij}$ separates the central cell $\Omega^b_{00}$ from its adjacent neighbor $\Omega^b_{ij}$, where the indices $(i,j)$ satisfy $i = 0, \pm1$ and $j = \pm(1 - |i|)$ (see Fig. 19.1 in  \citep{Buryachenko`2022}).
The representative BFUC $\Omega^b_{00}$ undergoes deformation in the same repetitive manner as all neighboring cells, ensuring periodicity of the body force:
\BB
\label{2.7}
{\color{black}\bfb(\bfx-\bfchi)=\bfb(\bfx),\ \ \ \bfchi\in {\bfLa}^b.}
\EE

This force distribution induces a displacement field given by ($\bfx\in\Omega_{00}^b$)
\BB
\label{2.8}
\bfu^{b(0)}(\bfx)\equiv - ({\bfcL}^{(0)})^{-1} \bfb.
\EE
Alternatively, this displacement can be represented via the periodic Green operator $\bfG^{(0)}(\bfx)$ corresponding to the Navier equation (\ref{2.4}) for the homogeneous reference tensor $\bfL^{(0)}$:
\BB
\label{2.9}
\bfu^{b(0)}(\bfx)= \int \bfG^{(0)}(\bfx-\bfy)\bfb(\bfy)~d\bfy.
\EE
Loosely speaking, the Green operator $\bfG^{(0)}(\bfx)$ may be interpreted as the inverse of the reference stiffness operator, characterizing the response of the infinite medium to localized force distributions. 

We now proceed to the consideration of a composite material (CM) whose local stiffness is given by $\bfL(\bfx)=\bfL^{(0)}+\bfL_1(\bfx)$, where $\bfL^{(0)}$ is the stiffness tensor of a homogeneous reference medium and $\bfL_1(\bfx)$ represents spatial variations.

The total displacement field $\bfu(\bfx)$ can then be decomposed as:
\BB
\label{2.10}
\bfu(\bfx)=\bfu^{b(0)}(\bfx)+\bfu_1(\bfx), 
\EE
where $\bfu^{b(0)}(\bfx)$ solves the homogeneous problem with the reference modulus:
\BBEQ
\label{2.11}
{\bfcL}^{(0)}(\bfu^{b(0)})
(\bfx)\! &=&\! -\bfb(\bfx),\\
\label{2.12}
{\bfcL}(\bfu)
(\bfx)
\! &=&\! {\bfcL}^{(0)}(\bfu^{b(0)}).
\EEEQ
Substituting these into the equilibrium equation (\ref{2.2}) yields:
\BB
\label{2.13}
\nabla\bfL^{(0)}\nabla \bfu_1(\bfx)=- \nabla\bfL_1(\bfx)\nabla\bfu(\bfx), 
\EE
which leads to an implicit integral representation of the strain field using a Green operator called a modified Lippmann-Schwinger
(ML-S) equation:
\BB
\label{2.14}
\bfep(\bfx) =\bfep^{b(0)}(\bfx)+\int\bfU^{(0)}(\bfx-\bfy)\bftau(\bfy)~d\bfy, 
\EE
where $\bfep^{b(0)}(\bfx) = \nabla^s \bfu^{b(0)}(\bfx)$ is the symmetric gradient of the reference displacement, and $\bftau(\bfy) := \bfL_1(\bfy)\bfep(\bfy)$ is the polarization tensor.
The kernel $\bfU^{(0)}$ is the second derivative of the reference Green tensor $\bfG^{(0)}$, i.e.,
$\bfG^{(0)}$: $U^{(0)}_{ijkl}(\bfx)
=\nabla_j\nabla_ lG^{(0)}_{(ij)(kl)}$ which vanishes as $|\bfx| \to \infty$.
Here, the symmetrization in the lower indices (denoted by parentheses) ensures the tensor satisfies symmetry requirements of elasticity.
{\color{black} It should be noted that the solutions to Eqs. (\ref{2.11}) and (\ref{2.12}), given in the form of the volume integral representations (\ref{2.8}) and (\ref{2.14}), respectively, can also be obtained by alternative methods, such as the finite element method (FEM)..}

The solution to the integral equation (\ref{2.14}) can be expressed in terms of a Neumann series expansion, which iteratively approximates the solution by a convergent sequence of successive operator applications. This representation is valid under conditions ensuring the contraction property of the associated integral operator—typically guaranteed when the contrast between material properties is sufficiently small or the norm of the perturbation operator is less than one. Formally, the Neumann series takes the form:
\BB
\label{2.15}
\bfep =\Big[\sum_{k=0}(\bfU^{(0)}\ast\bfL_1)^k\Big]\bfep^{b(0)}. 
\EE

The strain Eshelby-Green tensor $\bfU^{(0)}(\bfx)$ is known explicitly in the Fourier domain for an infinite, isotropic reference medium with Lamé coefficients $\lambda^{(0)}$ and $\mu^{(0)}$ (see, e.g.,  \citep{{Moulinec`S`1994},  {Mura`1987}}):
\BBEQ
\label{2.16}
{\bfU}^{(0)}&=&{1\over 4\mu^{(0)}}{\bfU}_1+
{\lambda^{(0)}+\mu^{(0)}\over \mu^{(0)}(\lambda^{(0)}+2\mu^{(0)})}{\bfU}_2, 
\\
\label{2.17}
{U}_{1|ijkl}(\bfze)&=&-|\zeta|^{-2}(\delta_{ki}\zeta_h\zeta_j+\delta_{hi}\zeta_k\zeta_j+\delta_{kj}\zeta_h\zeta_i+\delta_{hj}\zeta_k\zeta_i), \\
\label{2.18}
{U}_{2|ijkl}(\bfze)&=&|\zeta|^{-4}\zeta_i\zeta_j\zeta_k\zeta_h,% \over |\bfxi|^4}.
\EEEQ
It is also worth noting that for any square-integrable field $\bftau(\bfx)$, the convolution $({\bfU^{(0)}} * \bftau)(\mathbf{0}) = \mathbf{0}$, ensuring consistency with the equilibrium condition.

Equation (\ref{2.14}) is supplemented by the periodic boundary conditions (PBC) at the body force unit cell (BFUC) 
$\Omega_{00}^b$: 
\BB
\label{2.19}
\bfep(\bfx) \#, \ \ \ \bfsi(\bfx)\cdot \bfn(\bfx)=-\#, \ \ \bfx\in \partial \Omega^b_{00}
\EE
where the strain field $\bfep(\bfx)$ is assumed to be periodic (denoted by $\bfep \#$), while the corresponding traction vector $\bfsi(\bfx) \cdot \bfn(\bfx)$ is anti-periodic (denoted by $\bfsi\cdot \bfn=-\#$). These conditions ensure compatibility and equilibrium across the boundaries of adjacent periodic cells. The PBC (\ref{2.19}) at $\bfb(\bfx)\equiv {\bf 0}$ are equivalent to the homogeneous remote boundary conditions (also called the
kinematic uniform boundary conditions (KUBC) and static uniform boundary
conditions (SUBC), respectively) with some some symmetric constant tensors either
$\bfep^{w_{\Gamma}}$ or $\bfsi^{w_{\Gamma}}$ 
\BBEQ
%\label{2.20}
\bfu(\bfy)&=& \bfep^{w_{\Gamma}}\bfy, \ \forall\bfy\in \Gamma_{0u}={\Gamma}_0,\nonumber\\
\label{2.20}
\bft(\bfy)&=&\bfsi^{w_{\Gamma}}\bfn(\bfy), \ \forall\bfy\in \Gamma_{0\sigma}={\Gamma}_0,
\EEEQ
correspond
to the analyses of the equations for either strain or stresses, respectively, which are formally similar to each other.

It is further assumed that the composite material
$w=\cup \Omega_{ij}$ ($i,j=0,\pm 1, \pm 2,\ldots)$ exhibits periodic microstructure, with a unit cell (UC) of geometry $\Omega_{ij}$
 and a grid of centers  $\bfLa=\{\bfx_{ij}\}$ of $\Omega_{ij}$, and that the body force $\bfb(\bfx)$ is defined over an enlarged domain $\Omega^b_{00}$ such that representative UC $\Omega_{00} \subset \Omega^b_{00}$.:
\BB
\label{2.21}
{\color{black}\bfL(\bfx-\bfchi)=\bfL(\bfx),\ \ \ \bfchi\in {\bfLa}.}
\EE
Under these assumptions, the modified Lippmann-Schwinger equation (\ref{2.14}), along with the PBC (\ref{2.19}), remains valid over $\Omega^b_{00}$. However, the standard periodic boundary conditions on the smaller reference unite cell (UC) $\Omega_{00}$:
\BB
\label{2.22}
\bfep(\bfx) \#, \ \ \ \bfsi(\bfx)\cdot \bfn(\bfx)=-\#, \ \ \bfx\in \partial \Omega_{00}
\EE
are generally violated due to the larger support of the forcing term $\bfb(\bfx)$, which extends beyond $\Omega_{00}$. As a result, equilibrium between adjacent unit cells $\Omega_{ij}$ is not automatically satisfied, and the problem must be formulated on the extended domain $\Omega^b_{ij}$ to ensure consistency. 

{\color{black} We consider two distinct problems. The first involves Eq. (\ref{2.4}) subjected to BFCS (\ref{2.5}) on an isolated domain $\Omega_{00}^b$ with a free edge. The second involves Eq. (\ref{2.4}) in an infinite periodic medium subjected to the periodic body force (\ref{2.7}) with PBCs (\ref{2.19}) (or equivalently, remote BCs (\ref{2.20})). At the end of Section 5, we establish the conditions under which the solutions to these problems coincide within the isolated domain $\bfep(\bfx)\in \Omega_{00}^B$ and the representative domain  $\Omega_{00}^b$ of the periodic medium.}

We now consider the classical Lippmann-Schwinger problem in the absence of body forces, i.e., $\bfb\equiv {\bf 0}$ ($\bfx\in R^d$) ,
with periodic boundary conditions (PBC) applied on the boundary of the UC $\Omega_{00}$, as defined in Eq.~(\ref{2.22}).
In this setting, we adopt an alternative decomposition of the displacement field, replacing Eq.~(\ref{2.10}) with:
\BB
\label{2.23}
\bfu(\bfx)=\lle\bfep\rle\cdot\bfx+\bfu_1(\bfx), 
\EE
where $\langle \bfep \rangle$ is a prescribed constant macroscopic strain, and $\bfu_1(\bfx)$ is a periodic fluctuation field.
Substituting this decomposition into the governing equations yields the same differential identity as in Eq.~(\ref{2.13}), albeit with a different fluctuation field $\bfu_1(\bfx)$. This leads directly to the classical Lippmann-Schwinger (L-S, see  \citep{{Moulinec`S`1994},
 {Moulinec`S`1998}}) integral equation:
\BB
\label{2.24}
\bfep(\bfx) =\lle\bfep\rle+\int\bfU^{(0)}(\bfx-\bfy)\bftau(\bfy)~d\bfy. 
\EE
Here, $\bfU^{(0)}(\bfx)$ refers to the strain-based Green operator previously introduced in Eq. (\ref{2.15}) (see  \citep{{Buryachenko`2007}, {Buryachenko`2022}, {Mura`1987}}). It is worth noting that some works use ${\bf \Gamma}^{(0)} = -\bfU^{(0)}$ instead of $\bfU^{(0)}$ (see  \citep{{Moulinec`S`1994}, {Moulinec`S`1998}}). {\color{black}It is noteworthy that the ML-S equation (\ref{2.14}) is simpler than the classical L-S equation (\ref{2.24}) where its free term $\langle \boldsymbol{\varepsilon} \rangle$ (\ref{2.24}) is defined by the averaged solution. In contrast, the free term $\boldsymbol{\varepsilon}^b(\mathbf{x})$ in equation (\ref{2.14}) is a deterministic function that must be estimated in advance using equation (\ref{2.9}).}

In this formulation (\ref{2.24}), the total strain field $\bfep(\bfx)$ is periodic, and the macroscopic strain $\langle \bfep \rangle$ serves as the driving term. 
The solution to the integral equation (\ref{2.24}) can be represented using a Neumann series
\BB
\label{2.25}
\bfep =\Big[\sum_{k=0}(\bfU^{(0)}\ast\bfL_1)^k\Big]\lle\bfep\rle, 
\EE
the properties of which are detailed in  \citep{Milton`2002}. The operator representations on the right-hand sides of Eqs. (\ref{2.15}) and (\ref{2.25}) are identical and, consequently, are subject to the same conditions of convergence.

The method (\ref{2.25}) is equivalent to an explicit operator recurrence procedure, initiated from the expression
$\bfep^1=\lle\bfep\rle$:
\BB
\label{2.26}
\bfep^{[i+1]} =\lle\bfep\rle+\bfU^{(0)}\ast ( \bfL_1\bfep^{[i]}). 
\EE
 \citep{Michel`et`1999} observed that the strain Green's operator satisfies the identity:
\BB
\label{2.27}
\bfU^{(0)}\ast ( \bfL^{(0)}\bfep) =\lle\bfep\rle-\bfep.
\EE
Substiting Eq. (\ref{2.27}) into Eq. (\ref{2.26}) yields an equivalent recurrence relation in the form:
\BB
\label{2.28}
\bfep^{[i+1]} =\bfep^{[i]}+\bfU^{(0)}\ast ( \bfL\bfep^{[i]})\ \ {\rm with}\ \ \bfep^{[1]}=\lle\bfep\rle. 
\EE
We observe that the recurrence counterpart (\ref{2.28}) of the classical L-S equation Eq. (\ref{2.24}) is reduced to its equivalent recursive representation (\ref{2.28}) by applying Eq. (\ref{2.27}), under the condition that $\bfL_1(\bfx)\equiv \bfL^{(0)}$. 

We note that the modified Lippmann-Schwinger equation Eq. (\ref{2.14}) can be viewed as a generalization of the classical form Eq.(\ref{2.24}), in which the constant macroscopic strain $\langle \bfep \rangle$ is replaced by a non-uniform strain field $\bfep^{b(0)}(\bfx)$ that possesses compact support, mirroring the localized nature of the body force $\bfb(\bfx)$ introduced in Eq.~(\ref{2.5}).

\section{Classical and new RVE concepts}
\setcounter{equation}{0}
\renewcommand{\theequation}{3.\arabic{equation}}

The concept of the Representative Volume Element (RVE), originally introduced by \citep{Hill`1963}, has a long and complex history, often marked by debate and reinterpretation (for a detailed account, see  \citep{Ostoja`et`2016}). To faithfully convey the essence of Hill’s original definition and its foundational importance, we cite directly from Hill’s seminal work  \citep{Hill`1963}, which provides a rigorous basis for the RVE concept (at $\bfb(\bfx)\equiv {\bf 0}$).

\noindent{\bf Definition 2.1.} {\it Representative volume element (RVE)
(a) is structurally entirely typical of the whole mixture on average, and
(b) contains a sufficient number of inclusions for the apparent overall moduli to
be effectively independent of the surface values of traction and displacement, so
long as these values are ‘macroscopically uniform.... The contribution of this surface layer to any average can be negligible by taking the sample large enough.}

The primary role of the RVE is to provide an estimation of the effective moduli $\bfL^*$, which in turn relies on the evaluation of phase-averaged field quantities. For composite materials with periodic microstructures, the RVE coincides with the Unit Cell (RVE~$\equiv$UC), and the application of remote homogeneous boundary conditions (\ref{2.20})  naturally reduces to periodic boundary conditions (PBC, (\ref{2.22})). 
Although \citep{Hill`1963} originally formulated the RVE concept in the context of linearly locally elastic
CMs, his Definition 2.1 notably omits the term ``constitutive law," allowing for a natural extension of the concept to nonlinear and nonlocal material models.

{\color{black}A representative response can be obtained via DNS of microstructural volume elements (MVEs), either synthetically generated or experimentally extracted (e.g., by micro-CT). The homogenized overal properties $\bfL^{\rm A}_{\rm KUBC}$ and $\bfM^{\rm A}_{\rm SUBC}$ are computed under kinematically (\ref{2.20}$_1$) and statically (\ref{2.20}$_2$)  uniform boundary conditions, respectively. While generally different, their discrepancy
\BB
\label{3.1}
\bfL^{\rm A}_{\rm KUBC}-(\bfM^{\rm A}_{\rm SUBC})^{-1}\to {\bf 0},
\EE
decreases with increasing MVE size, enabling estimation of the RVE size and the effective moduli $\bfL^*$. The RVE concept assumes sufficiently large—ideally infinite—material domains (see, e.g., \cite{Ostoja`et`2016}; \cite{Buryachenko`2022}, p. 226). 
It should be emphasized that, although increasing the number of realizations in this intuitive EVE framework can achieve statistical convergence in estimating $\bfL^*$ (i.e., the mean sample response), it does not inherently remove edge-related scale effects. 
In practice, intuitive RVEs (``sufficiently large sample") are typically constructed using two main approaches: (1) simulated random inclusion fields and (2) image-based models derived from micro-CT scans of real materials (see, e.g., \citep{{Konig`et`1991},  {Ohser`M`2000}, {Torquato`2002}}).
A pictorial interpretation of the RVE concept is illustrated in Fig. 1, which shows a micro-CT scan of a heterogeneous material subjected to remote homogeneous loading, as defined by Eq. (\ref{2.20}$_1$).}

Under the application of PBC (\ref{2.22}) (or (\ref{2.20}$_1$), the estimation of effective moduli $\bfL^*$ (the "upscaling" procedure) and the determination of field concentration factors $\bfA(\bfz)$ for points $\bfx \in v_i$ (the ``downscaling" procedure) are expressed as:

\vspace{1.mm} \noindent \hspace{30mm}
\parbox{8.8cm}{`
\centering \epsfig{figure=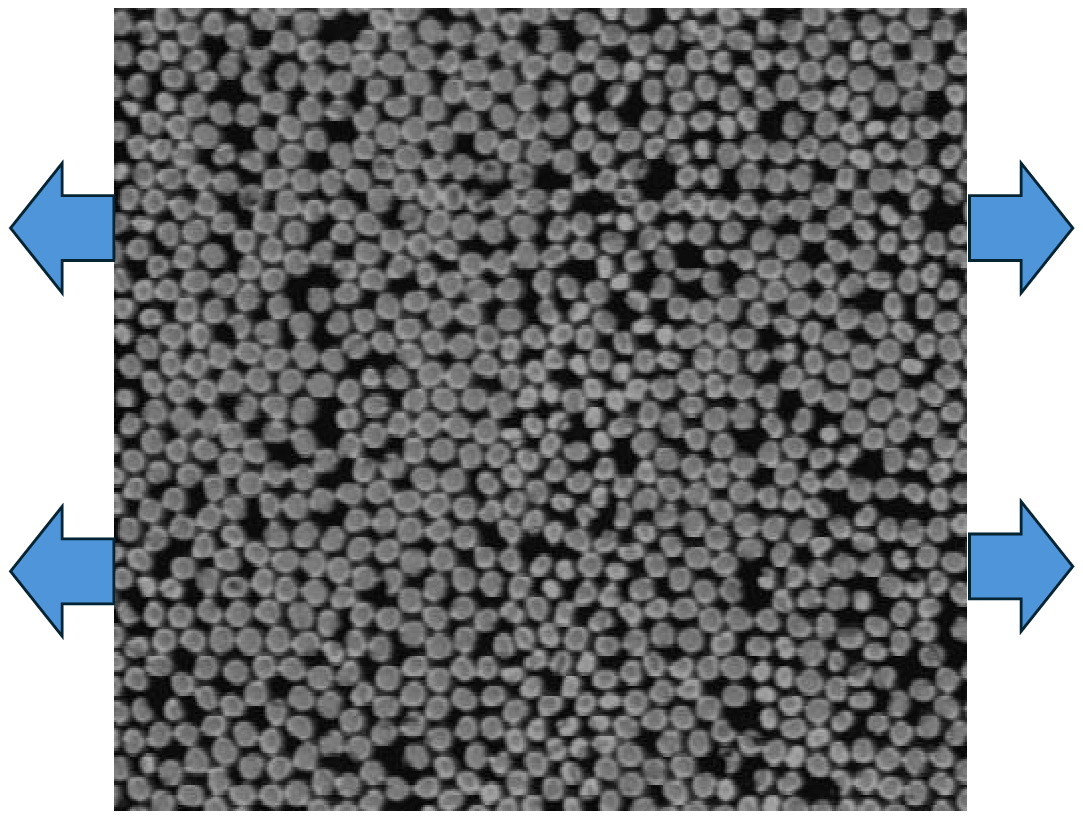, width=7.8cm}\\ \vspace{-50.mm}
\vspace{122.mm}
\vspace{-117.mm} \tenrm \baselineskip=8pt
{\color{black} { Fig. 1:} CT image with remote BC (\ref{2.20}) }}
%(\ref{2.4})}}
%(\ref{2.20}$_1$) }}
\vspace{2.mm}

\BBEQ
%\label{3.1}
\!\bfL^*=\bfL^{(0)}+\bfR^*, \ \ \lle\bftau\rle=\bfR^*\lle\bfep\rle,\ \ 
\label{3.2}
\lle\bfep\rle_i(\bfx)=\bfA^*(\bfx)\lle \bfep\rle
\EEEQ
highlighting the intrinsic coupling between upscaling and downscaling—two complementary aspects of the same homogenization framework.

Under the PBC as defined in Eq. (\ref{2.22}), it follows from Eqs. (\ref{3.2}) that both the effective moduli $\bfL^*$ and the effective strain field $\langle \bfep \rangle_i(\bfx)$ are independent of the particular choice of UC $\Omega_{ij}$. However, this invariance breaks down in the presence of a body force applied to the special BFUC $\Omega_{00}^b$, where the periodic boundary condition at $\Omega_{00}$ (\ref{2.22}) is no longer satisfied. In this case, the resulting strain field $\bfep(\bfX)$, expressed in the coordinate system associated with the body force $\bfb(\bfX)$ (with $\bfX \in \Omega_{00}^b$), acquires a dependence on the specific unit cell $\Omega_{ij}^b$ containing the point $\bfX$.

To eliminate this dependency and restore generality, we analyze the behavior of periodic composite materials (CMs) under a fixed body force, while allowing the microstructure (i.e., the grid of unit cell centers $\bfx_{\bf m} \in \bfLa$) to undergo a rigid translation. For concreteness, we restrict the discussion to the linear elasticity model defined by Eqs. (\ref{2.2}), and the microstructural strain solution is denoted by $\bfep_0(\bfx)$. When the grid of unit cell centers undergoes a rigid translation by a vector $\bfchi$ (i.e., $\bfLa_0 \to \bfLa_{\bf\chi}$), the material stiffness $\bfL(\bfx, \bfchi)$ (\ref{2.2}), the phase indicator function $V_i(\bfx, \bfchi)$, and the resulting strain field solution $\bfep(\bfx, \bfchi)$ governed by Eqs. (\ref{2.1})–(\ref{2.3}) will also transform accordingly
\BBEQ
\label{3.3}
\bfL(\bfx,\bfchi)&=&\bfL_0(\bfx-\bfchi), \ \ V_i(\bfx,\bfchi)=V_{i0}(\bfx-\bfchi),\\
\bfb(\bfx,\bfchi)&=&\bfb_0(\bfx),\ \ \bfu(\bfx,\bfchi)\not\equiv \bfu_0(\bfx-\bfchi).
\label{3.4}
\EEEQ
The inequality in Eq.~(\ref{3.4}$_2$) holds because the body force field $\bfb(\bfx)$ is fixed.

Accordingly, for every translation vector $\bfchi \in \cV_{\bf x}$, the solution $\bfu(\bfx, \bfchi)$ derived from Eqs.~(\ref{3.4})  enables the definition of an effective (macroscopic) displacement field over the domain $w$.
\BBEQ
\label{3.5}
\!\!\!\!\!\!\!\!\!\!\!\!\!\!\!\!\!\!\!\!\!\!\lle\bfu\rle(\bfx)\!&=&\!{1\over\overline
{\cV}_{\bf x}}\int_{{\cal V}_{\rm \bf x}}\bfu(\bfx,\bfchi)~d\bfchi, \
\lle\bfu\rle^{l(1)}(\bfx)= {1\over\overline{\cV}_{\bf x}}\int_{{\cal V}_{\rm \bf x}}\bfu(\bfx,\bfchi)V_i(\bfx,\bfchi)~d\bfchi, \\
\label{3.6}
\!\!\!\!\!\!\!\!\!\!\!\!\!\!\!\!\!\!\!\!\!\!\!\!\lle\bfsi\rle(\bfx)\!&=&\! {1\over\overline
{\cV}_{\bf x}}\int_{{\cal V}_{\rm \bf x}}\bfsi(\bfx,\bfchi)~d\bfchi,\
\lle\bfsi\rle^{(1)}(\bfx)= {1\over\overline
{{\cal V}}_{\bf x}}\int_{\cV_{\rm \bf x}}\bfsi(\bfx,\bfchi)V_i(\bfx,\bfchi)~d\bfchi,
\EEEQ
It is interesting that Eqs. (\ref{3.5}$_1$) and (\ref{3.6}$_1$) formally looks as averaging over the moving-window ${\cV}_{\bf x}$ although the operations (\ref{3.5}$_1$) and (\ref{3.6}$_1$) are conceptually different and obtained by averaging over the number of the displacement realizations $\bfu(\bfx,\bfchi)$ produced by a parallel transform of $\bfL(\bfx,\bfchi)$ (\ref{3.4}) rather than by averaging of one realization
$\bfu_0(\bfx)$ over the moving-window.
The averages defined in Eqs.~(\ref{3.5}) and (\ref{3.6}) are more accurately described as ensemble averages, taken over the statistical set of all translated configurations of a given periodic microstructure, where the translations $\bfchi$ are uniformly distributed within the periodicity cell $\cV_{\bf x}$. This {\it translation-based averaging} procedure is general and applies to periodic composite materials with arbitrary constitutive behavior and under any inhomogeneous loading conditions. {\color{black} In so doing, the solution $\{\cdot\}(\bfx,\bfchi)\}$ for each $\bfchi \in \cV_{\bf x}$ can be obtained by any sutable numerical method (e.g., FEA or FFT, see section 5)}.
A particular instance of such an averaging scheme—specifically, in the form of Eq.~(\ref{3.6}$_1$)—was introduced in the context of asymptotic homogenization by  \citep{Smyshlyaev`C`2000} and later discussed by  \citep{Ameen`et`2018}. Notably,  \citep{Smyshlyaev`C`2000} attributes this idea to a personal communication with J.R. Willis, in the context of periodic media subjected to periodic loading.
Interestingly, the statistical averages in (\ref{3.5}$_2$), and (\ref{3.6}$_2$) can be interpreted as a probabilistic reformulation of the classical student problem: ``What is the probability that a randomly dropped coin (representing an inclusion $v_i$) lands on a fixed point $\bfx \in {R}^d$?” In the specific cases of Eqs.(\ref{3.5}$_2$) and (\ref{3.6}$2$), the ``coin" (i.e., inclusion $v_i$) belongs to a periodically structured grid $\bfLa_{\bf\chi}$ undergoing random translations, further emphasizing the ensemble nature of the averaging.
An illustration of the averaging scheme described by Eq.~(\ref{3.6}) for the one-dimensional case is shown in Fig.~2. In this 

\vspace{2.mm} \noindent \hspace{10mm} \parbox{6.2cm}{
\centering \epsfig{figure=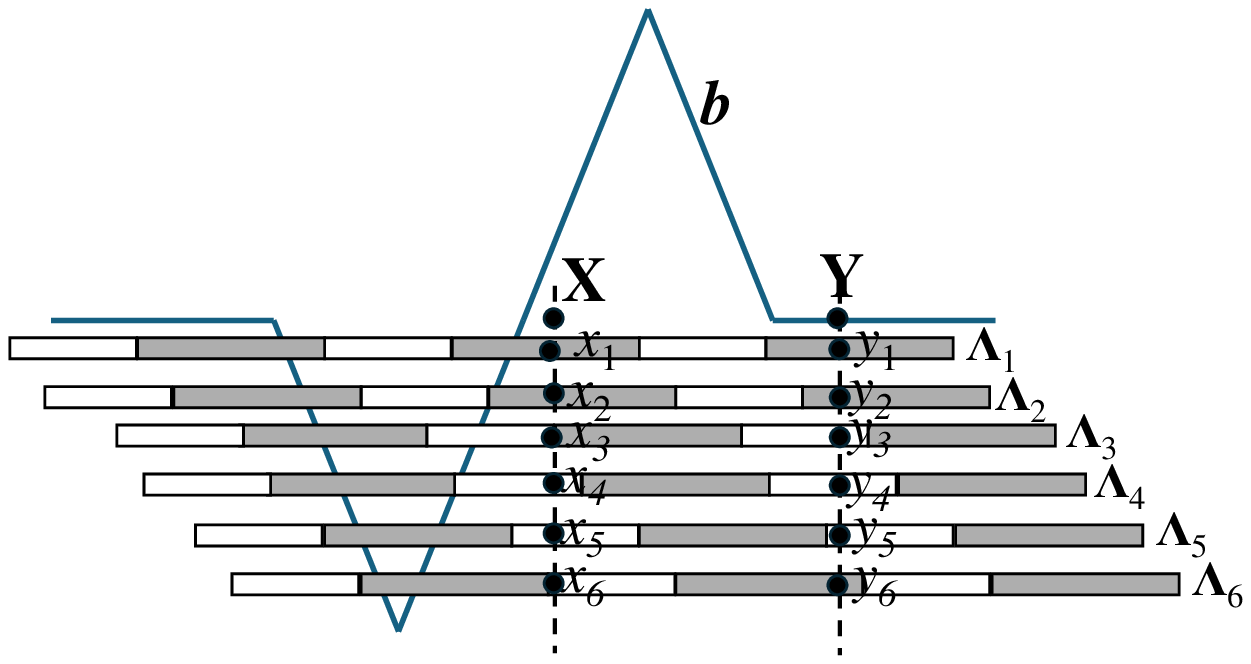, width=10.2cm}\\ \vspace{-22.mm}
\vspace{20.mm}
\vspace{-72.mm} \tenrm \baselineskip=8pt
{{\sc Fig. 2:} Scheme of translated averaging}}
\vspace{2.mm}

\noindent example, a fixed macroscopic point 
%\noindent 
$\bfX \in b({\bf0}, B^b)$ corresponds to a set of local points $\bfx_j = \bfX - \bfchi$ located within the translated unit cell grids $\bfLa_{\bf\chi} = \bfLa_j$ for $j = 1,\ldots,6$. The estimation of the average stress field $\langle \bfsi \rangle(\bfX)$ is carried out by summing the stress values $\bfsi(\bfX, \bfchi) = \bfsi(\bfx_j)$ across all realizations of the translated grid $\bfLa_j$.
In contrast, the average stress within the inclusions, denoted by $\langle \bfsi \rangle^{(1)}(\bfX)$, is computed by summing only over those realizations where the point $\bfX$ falls inside an inclusion—specifically, the realizations corresponding to $j = 1, 2, 3, 6$.
Notably, due to nonlocal effects, it is possible to observe non-zero average stress values $\langle \bfsi \rangle(\bfY) \ne \mathbf{0}$ at points $\bfY \notin b({\bf0}, B^b)$, even though the body force vanishes at those points, i.e., $\bfb(\bfY) = \mathbf{0}$. This nonlocal average $\langle \bfsi \rangle(\bfY)$ is again computed by summing $\bfsi(\bfX, \bfchi) = \bfsi(\bfx_j)$ over all six realizations ($j = 1,\ldots,6$), while the inclusion-specific average $\langle \bfsi \rangle^{(1)}(\bfY)$ is based only on realizations $j = 1, 2, 6$, where $\bfY$ lies within an inclusion.

After the avaraging (\ref{3.5}) and (\ref{3.6}) over the grid translation $\bfLa{\bf \chi}$, we can introduce a dataset $\bfcD^{\rm p}$ for periodic-structured CMs as follows:
\BBEQ
\!\!\!\!\!\!\!\!\!\!\!\!\!{\bfcD}^{\rm p}&=&\{\bfcD^{\rm p}_k\}^N_{k=1}, \ \ \bfcD^{\rm p}_k=\{\lle{\bfu}_k\rle(\bfb_k,\bfx),\!\lle\bfsi_k\rle(\bfb_k,\bfx),
\nonumber\\
\label{3.7}
&&\!\lle{\bfep}_{k}\rle^{(1)}(\bfb_k,\bfx),
\!\lle{\bfsi}_{k}\rle^{(1)}(\bfb_k,\bfx), \bfb_k(\bfx)
\},
\EEEQ
where each set of effective parameters is computed for a given body force realization $\bfb_k$ using the averaging procedures defined in Eqs.~(\ref{3.5}) and (\ref{3.6}) during the offline stage. The coordinates $\bfx \in \mathbb{R}^d$ represent macroscopic
spatial variables; hence, the dataset $\bfcD^{\rm p}$ provides macroscopic-level information only. 
Although each sub-dataset $\bfcD^{\rm p}_{{\bf \chi}k}$, corresponding to a specific grid $\bfLa{\bf\chi}$, is generated through DNS,
the complete dataset $\bfcD^{\rm p}_k$ is obtained using the translated averaging procedures defined in Eqs. (\ref{3.5}) and (\ref{3.6}). This averaging technique is a central component of computational analytical micromechanics (CAM), as defined in  \citep{{Buryachenko`2024},  {Buryachenko`2024b}}). Furthermore, the datasets $\bfcD^{\rm r}$ for random-structure composite materials, also derived  by \citep{Buryachenko`2024b} using CAM, are formally equivalent to the 
periodic dataset $\bfcD^{\rm p}$ 
given in Eq. (\ref{3.7}). As a result, the estimation procedure for 
$\bfcD^{\rm p}$ in Eq.~(\ref{3.7}) is likewise referred to as the CAM method.
We use the displacement field
$\bfu(\bfx)$ in Eq. (\ref{3.7}) (see also Eq. (\ref{3.8})),
only because the source publications for Section 6 are given as $(\bfu(\bfx),\bfb(\bfx))$ rather than the more typical
$(\bfep(\bfx),\bfb(\bfx))$; absent this data restriction, one could equally replace $\bfu(\bfx)$ with the strain field $\bfep(\bfx)$. 

The representation for the datased $\bfcD^{\rm p}$ (\ref{3.7}) was obtained for either a finite seze body force UC $\Omega_{00}^b$ or 
a full size $\Omega_{00}^b=R^d$ with finite size compact support ${\cal B}$ (\ref{2.5}) (i.e. $B^b<\infty$). 
Reformulation and generalization of the classical definition  by \citep{Hill`1963} 2.1
enables one to formulate a flexible definition sufficient for our current interests with self-equilibrated body force $\bfb(\bfx)$ (\ref{2.5}):

\noindent {\bf Definition 2.2.} {\it RVE is structurally entirely typical of the whole CM area, which is sufficient for all apparent effective parameters 
$\bfcD^{\rm p} _k$ ($k=1,\dots,N)$
(\ref{3.7}) to be effectively stabilized outside $\bfx\not\in {\rm RVE}$ (i.e. the strains and stresses vanish in 
$\bfx\in \overline {\rm RVE}:=\Omega^b_{00}\setminus {\rm RVE})$ 
in the infinite periodic structure CMs.}

A critical issue in the analysis of periodic CMs is the appropriate choice of PBC (\ref{2.22}) a at the interface of UCs provide a connection to field distributions, which is the neighboring UCs. PBC (\ref{2.22}) is strictly valid under homogeneous remote loading conditions (\ref{2.20})and zero body force $\bfb({\bfx})\equiv {\bf 0}$ (\ref{2.5}).
However, in the more general case where a nonzero body force field is present, the standard PBC given by Eq.~(\ref{2.22}) becomes inaccurate and may not reflect the true mechanical response of the material. In such scenarios, it is important to note that imposing PBC at the unit cell level is not strictly necessary, especially when direct numerical simulation (DNS) is employed. 
In particular, if the estimates in Eqs. (\ref{2.22}) is eliminated, the size of the RVE in this context becomes a tunable parameter, which must be determined to satisfy a prescribed tolerance criterion ($|\bfsi(\bfx)|, |\bfep(\bfx)|<{\rm tol}$, \ $\bfx\in \overline{\rm RVR}$) ensuring that the geometrical and mechanical representativeness of the RVE is sufficient for accurate homogenization.
Moreover, owing to the compact support of the applied body forces $\{\bfb_k(\bfx)\}_{k=1}^N$, the original problem posed on the infinite domain $\mathbb{R}^d$ for periodic CM is effectively reduced to a finite domain corresponding to the RVE. This reduction significantly simplifies the analysis. As a result, complications typically associated with finite sample size or boundary-induced (edge) effects are inherently avoided.

{\color{black} Definition 2.2 is conceptually distinct from Definition 2.1. The latter pertains to an idealized, asymptotically large (infinite) domain, where effective material parameters are obtained through a formal limiting process for a “sufficiently large” sample. In contrast, Definition 2.2 introduces a more practical concept—a finite-size RVE—serving as an initial approximation that is amenable to computational or experimental implementation and subsequent refinement. The key distinction, however, lies in the fundamentally different remote BCs (\ref{2.22}) and BFCS (\ref{2.6}) employed in Definitions 2.1 and 2.2, respectively.}
This definition considers a heterogeneous medium periodically structured occupying the entire space $\mathbb{R}^d$. Similarly, no reference is made to an explicit ``effective moduli" (or ``effective nonlocal operator," see for references  \citep{Buryachenko`2024}). Rather, attention is focused on the exterior domain $\overline{\text{RVE}} = \mathbb{R}^d \setminus \text{RVE}$, where dataset $\{\bfcD_k^p(\bfx)\}_{k=1}^N$ (\ref{3.7}) reach a stabilized state.
Stabilization of the dataset $\bfcD^p_k$ (\ref{3.7}) along with the proper selection of the RVE implies that all effective parameters within the annular region bounded by $|\bfx| = B^{\rm RVE}$ and $|\bfx| = B^{\rm RVE} + B^b/2$ remain constant within a prescribed tolerance. When this condition is met, the external region $|\bfx| > B^{\rm RVE} + B^b/2$ can be excluded from the simulation, allowing the infinite medium to be accurately modeled using a finite-sized sample. In this way, a correctly chosen RVE eliminates edge effects —commonly encountered as boundary-layer artifacts, especially when $\langle \bfep \rangle^\Omega(\bfx)$ remains nonzero near the domain boundary. This phenomenon is discussed, for example, on p. 129 of  \citep{Buryachenko`2007}.
Conversely, if the RVE is not appropriately selected (i.e., $B^{\rm RVE}$ is insufficiently large), the subsequent application of the dataset $\bfcD$ (\ref{3.7}) (as discussed in Section 6) will result in numerical errors originating from both finite sample size and residual edge effects.

Figure 3 illustrates the spatial arrangement $\bfcB^b\subset {\rm RVE}\subset\Omega^b_{00}$ for composite materials (CMs) with two types of microstructures: periodic (Fig. 3a) and deterministic (Fig. 3b). In Fig. 3a, we depict a representative periodic
%\noindent 
configuration $X$ of inclusions, characterized by the set of centers $\bfLa_{\bf \chi}$. Although periodicity underlies the inclusion layout, it is not directly used in simulations. Instead, datasets $\bfcD^{{\rm p}j}_{{\bf \chi}k}$ are generated by applying various realizations of BFCS loading (2.6), and evaluating the local response within the BFUC $\Omega^b_{00}$ using a suitable numerical method (e.g., 
{\color{black} FEA} of FFT-based solvers from Section 5). 
Importantly, stress and strain fields are assumed to vanish in the exterior of the RVE ($\overline{\mathrm{RVE}}$), allowing the treatment of a finite set of inclusions $v_i \subset \Omega^b_{00}$ without explicitly enforcing periodicity. The periodic structure $\bfLa_{\bf \chi}$ is leveraged only during the translation averaging process (see Eqs. (\ref{3.5}) and (\ref{3.6})).
In a similar manner, Fig. 3b represents deterministic 

\vspace{-1.mm} \noindent \hspace{-25mm} \parbox{16.2cm}{
\centering \epsfig{figure=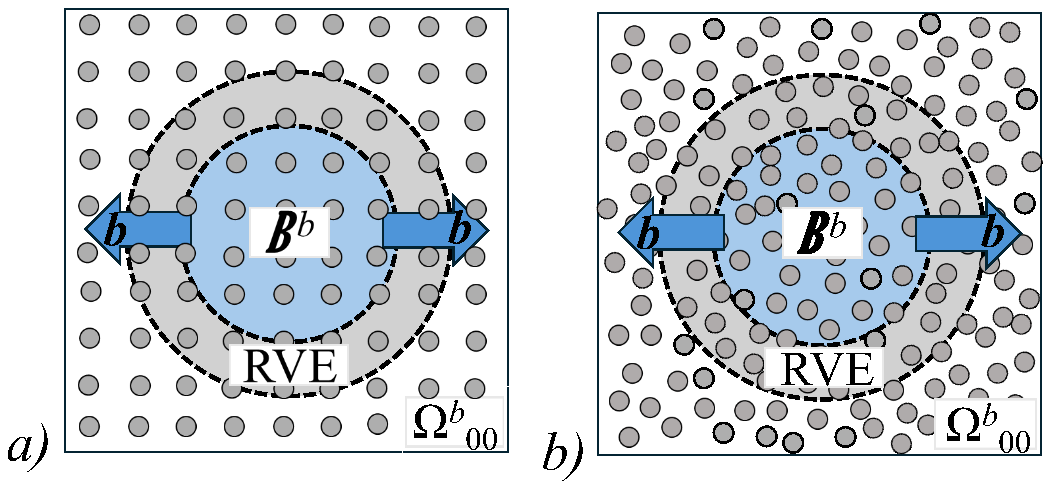, width=11.2cm}\\ \vspace{-22.mm}
\vspace{-72.mm} \tenrm \baselineskip=8pt
\centering{{\sc Fig. 3:} Schemes of $\bfcB^b\subset {\rm RVE}\subset\Omega^b_{00}$ for CM with periodic a) \\
and deterministic b) structures}}
\vspace{1.mm}

\noindent  microstructures $X^j$ ($j = 1, 2, \ldots$), which are neither periodic nor stochastic and may, for instance, correspond to distinct samples obtained from CT imaging. As in the periodic case, datasets $\bfcD^{\mathrm{d}j}_k$ are computed within $\Omega^b_{00}$ for different deterministic configurations $X^j$ and fixed body force fields $\bfb_k(\bfx)$. These yield an ensemble-based dataset defined (as (\ref{3.7})) by
\BBEQ
\label{3.8}
\!\!\!\!\!\!\!\!\!\!\!{\bfcD}^{\rm d}&\!=\!& \{\bfcD^{\rm d}_k\}_{k=1}^N,\ \ {\bfcD}^{\rm d}_k=\{\lle{\bfu}_k\rle(\bfb_k,\bfx),\lle\bfsi_k\rle(\bfb_k,\bfx), \nonumber\\
\!\!\!\!\!\!&&\!\!\!\!\!\!\!\lle{\bfep}_{k}\rle^{(1)}(\bfb_k,\bfx),
\!\lle{\bfsi}_{k}\rle^{(1)}(\bfb_k,\bfx), \bfb_k(\bfx)
\},
\EEEQ
where statistical averaging $\lle \cdot \rle$ is performed over the set of deterministic configurations $X^j$, capturing the effective material response to each BFCS $\bfcB^b$ (\ref{2.5}) realization.
No specific assumptions are made about the geometry of the regions $\bfcB^b$ or the RVE itself (cf. Definition 2.2), nor about their relative sizes. Spherical shapes shown in Fig. 3 are adopted for clarity and visualization purposes only, without implying any essential geometric constraint.

A particularly significant contribution to computational micromechanics is presented in the recent study by \citep{Silling`et`2024}, which proposes a novel framework for analyzing two-dimensional random composites. The authors examine a composite microstructure generated via Monte Carlo simulations within a finite square domain $w$, where each phase is modeled using nonlocal elasticity theory. The studied sample, containing approximately 900 circular inclusions (see Fig. 4a)
would conventionally be considered an ideal 
 candidate for implementing the classical RVE concept, as defined in Definition 2.1. 
In a notable departure from this traditional approach, the authors introduce a self-equilibrated body force with compact support, $\bfb(\bfx)$, as described in Eq.~(\ref{2.5}). This loading scheme is consistent with the methodologies developed in 
\citep{{Buryachenko`2023},  {Buryachenko`2023a}}, and fulfills the criteria outlined in the generalized RVE concept of Definition 2.2. Specifically, the distance between the RVE boundary and the outer boundary of the computational domain satisfies ${\rm dist}(\partial \mathrm{RVE}, \partial w) \approx 10a = 100\lambda$, where $\lambda$ is the characteristic spacing of the microstructural lattice.
The effectiveness of this approach is substantiated by the nearly vanishing strain field $\bfep(\bfx)$ observed near the external boundary $\partial w$, as shown in the violaceous region in Fig.~4b. This outer region corresponds to $\overline{\mathrm{RVE}} = w \setminus \mathrm{RVE}$, where strain localization is negligible.

\vspace{2.mm}% \noindent
%\vspace{-01.mm}% \noindent
\hspace{-10mm} \parbox{12.8cm}{%\hspace{-10mm}
\centering \epsfig{figure=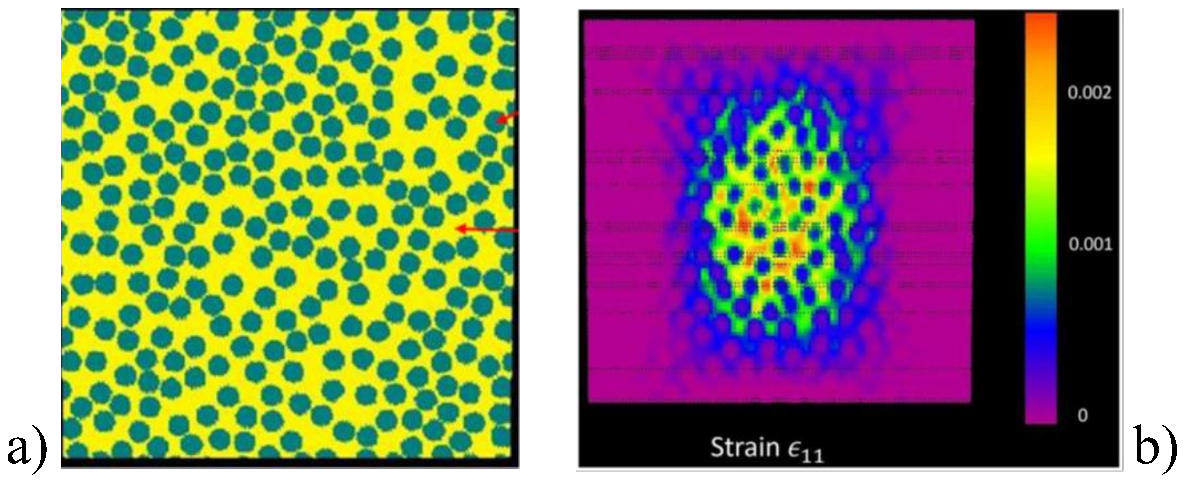, width=10.9cm}\\ \vspace{-22.mm}
\vspace{12.mm}
\vspace{-84.mm} \tenrm \baselineskip=8pt
{\hspace{5.mm}{\sc Fig. 4:} a) Simulated structure in $w$. b) DNS of strains $\bfep(\bfx)$ in $w$}}
\vspace{1.mm}

\noindent
Although the explicit term ``RVE” is not used in  \citep{Silling`et`2024}, the core conceptual structure clearly aligns with the generalized RVE framework, as represented by the central multicolored (non-violaceous) domain, which functions as the effective RVE.
{\color {black} To the best of the author’s knowledge, Fig. 4a provides the clearest illustration of the new RVE concept.}.
The ability of the RVE concept in Definition 2.2 to unify previously established methods and modeling approaches is of particular importance. This unifying capability not only confirms the internal consistency of the generalized definition but also demonstrates its flexibility and wide applicability. Such integrative potential is rare in micromechanics and represents a major step forward in establishing a more comprehensive and adaptable modeling paradigm for complex composite systems.

Due to the compact support of the applied body forces $\{\bfb_k(\bfx)\}_{k=1}^N$, the problem originally posed over the infinite domain $\mathbb{R}^d$ for both periodic and deterministiv structure CMs can, in practice, be reduced to a finite computational domain corresponding to the RVE. This localization greatly simplifies the analysis by removing the necessity of simulating the full-space domain. As a result, common issues related to limited sample size or edge effects are naturally avoided.
This type of problem can be tackled using a variety of computational mechanics methods, including the {\color{black} FEM, BEM, and others}. Notably, Moulinec and Suquet~ \citep{{Moulinec`S`1994},  {Moulinec`S`1998}} introduced an alternative and highly efficient technique based on fast Fourier transforms (FFTs)—commonly referred to as “FFT-based methods”—to solve computational homogenization problems of the type given in L-S equation (\ref{2.24}) {\color{black} with PBC (\ref{2.22}) at UC $\Omega_{00}$}. Their approach, known as the basic scheme, represented a major advancement at the time, as prior micromechanical simulations had been dominated by FEM-based techniques.
Thanks to the advantages discussed in the Introduction, FFT-based micromechanical methods now serve as competitive alternatives to FEM in a broad range of applications, including linear and nonlinear homogenization, local and nonlocal constitutive modeling, dislocation mechanics, and multiscale simulations. 
However, the solution of the modified L-S equation  (\ref{2.14}) {\color{black} with PBC (\ref{2.19}) at BFUC $\Omega_{00}^b$}   remains less explored. In this case, due to the periodicity requirement, the body-force unit cells $\Omega^b_{ij}$ (with $i,j = 0, \pm 1, \pm 2, \ldots$) are assumed to be finite in extent, i.e., $\Omega^b_{00} \neq \mathbb{R}^d$. 
Nevertheless, due to the vanishing of the fields in the exterior region $\overline{\rm RVE}$, extending the domain $\Omega^b_{00}$ does not affect the solution within the RVE—even in the limit where $\Omega^b_{00} \to \mathbb{R}^d$.
This approach enables solving Eq.   (\ref{2.14}) using the same FFT-based method for both periodic structures ($\Omega_{00} \ne \mathbb{R}^d$) and deterministic structures ($\Omega^b_{00} = \mathbb{R}^d$).
 In the next section, we focus on FFT-based methods for solving Eq.~(\ref{2.14}) under the assumption that the body force $\bfb(\bfx)$ (\ref{2.5}) has compact support.

\section{Discrete Fourier Transform}
\setcounter{equation}{0}
\renewcommand{\theequation}{4.\arabic{equation}}

In this section, we provide a brief overview of the notation and key properties of the Discrete Fourier Transform (DFT), tailored for direct use in the subsequent analysis.

We consider a unit cell $\Omega_{00}^b = \prod_{\alpha=1}^d [0, l^{\Omega}_{\alpha}]$, discretized using a uniform grid consisting of $N_1 \times \ldots \times N_d$ nodes along each spatial direction $\alpha = 1, \ldots, d$ 
\BBEQ
\label{4.1}
\bfx^{\bf k}_{\bf N}=\bfk\cdot\bfh\equiv[k_1 h_1,\ldots,k_d h_d]^{\top}, \ \ \bfk\in Z^d_{\bf N}:=\Big\{\bfm\in Z^d\Big|
0\leq m_{\alpha}\leq N_{\alpha}-1\Big\}.
\EEEQ
Hereafter, $\mathbb{Z}^d_{\bf N}$ denotes the set of $d$-tuples of integers corresponding to the grid indices, where each $N_{\alpha}$ ($\alpha = 1, \ldots, d$) is assumed to be even. We define the vector $\bfN := (N_1, \ldots, N_d)^{\top}$, with total number of grid points given by $|\bfN| = \prod_{\alpha=1}^{d} N_{\alpha}$. The grid spacing in the $\alpha$-th direction is $h_{\alpha} = l^{\Omega}_{\alpha} / N_{\alpha}$, and $\bfe_{\alpha}$ denotes the unit vector in the same direction.
A function $\bff:R^d\to R^d$ is said to be $\Omega_{00}^b$-periodic if it satisfies:
\BBEQ
\label{4.2}
\bff\Big(\bfx+\sum_{\alpha}\l^{\Omega}_{\alpha}k_{\alpha}\bfe_{\alpha}\Big)=\bff(\bfx),
\EEEQ
for
$\bfx\in \Omega_{00}$ and $\bfk\in Z^d$. In this section, we adopt the notation $\Omega_{00}^b = \prod_{\alpha}[0, l^{\Omega}_{\alpha}]$, which is commonly used in the context of the Discrete Fourier Transform (DFT) and its associated indexing. This contrasts with the earlier convention $\Omega_{00}^b = \prod_{\alpha}[-l^{\Omega}_{\alpha}, l^{\Omega}_{\alpha}]$ employed in Section 3.

If the integrable function $\bff$ satisfies the periodicity condition (\ref{4.2}), then its Fourier transform has a discrete frequency spectrum. The Discrete Fourier Transform (DFT) provides a means to transition between the spatial domain $\Omega_{00}^b$ and the corresponding frequency domain $\mathcal{F}_d$.
The DFT $\widehat{\bff} = \mathcal{F}_d(\bff)$ and its inverse (iDFT) $\bff = \mathcal{F}_d^{-1}(\widehat{\bff})$ for discrete periodic functions are defined as follows (see, e.g.,  \citep{Amidror`2013},  \citep{Marks`2009}):
\BBEQ
\label{4.3}
\widehat \bff_{\bf N}(\bfk)&=&\sum_{{\bf n}\in Z^d_{\bf N}} \bff_{\bf N}(\bfn) %\bfx^{\bf m}_{\bf N})
\omega^{-{\bf k}{\bf n}}_{\bf N},\\
\label{4.4}
\bff_{\bf N}(\bfn )%\bfx^{\bf k}_{\bf N})
&=& \frac{1} {|\bfN|} \sum_{{\bf k}\in Z^d_{\bf N}} \widehat \bff_{\bf N}(\bf k)
\omega^{{\bf k}{\bf n}}_{\bf N}
\EEEQ
where $\bfk, \bfn \in \mathbb{Z}^d_{\bf N}$ represent multi-index vectors associated with the $\bfN$-point discretization, and $\bff_{\bf N}(\bfn)$ and $\widehat{\bff}_{\bf N}(\bfk)$ are the corresponding sampled sequences (which may be complex-valued).
The Fourier kernel $\omega^{\bfk \cdot \bfn}_{\bf N}$ is given by:
\BBEQ
\label{4.5}
\omega^{{\bf k}{\bf n}}_{\bf N}={\rm exp}\Big(2\pi i\sum_{\alpha}\frac{k_{\alpha}n_{\alpha}}{N_{\alpha}}\Big),
\EEEQ
with $i = \sqrt{-1}$ and $\alpha$ indexing the spatial directions.
To emphasize the link between discrete and continuous Fourier representations, we interpret $\bff_{\bf N}(\bfn)$ and $\widehat{\bff}{\bf N}(\bfk)$ as sampled versions of the continuous functions $\bff(\bfx)$ and $\widehat{\bff}(\boldsymbol{\zeta})$. This interpretation allows the introduction of explicit spatial sampling steps $\Delta \bfh$ and frequency intervals $\Delta \boldsymbol{\zeta}$, such that values may be written as $\bff(\bfx^{\bfn}_{\bf N})$ and $\widehat{\bff}(\boldsymbol{\zeta}^{\bfk}_{\bf N})$ with $\boldsymbol{\zeta}^{\bfk}_{\bf N} = \bfk \cdot \Delta \boldsymbol{\zeta}$.
This approach has the advantage of associating each sample with its actual position in physical or frequency space, rather than a purely integer-based index. For discussions on various DFT indexing conventions, their relationships, and reasons for choosing one over another, see, for example,  \citep{Briggs`H`1995}. Nonetheless, when formulating DFT-related theorems, the sampled coordinates are typically abstracted to integer indices $\bfn$ and $\bfk$ for simplicity (see  \citep{Brigham`1988}).

The key properties of the DFT that are particularly relevant for the subsequent analysis include:
\BBEQ
\label{4.6}
\widehat\bff_{\bf N}(\bfk+\bfN)&=& \widehat\bff_{\bf N}(\bfk),\\
\label{4.7}
%\widehat\bff_{\bf N}(\bfk)&=& \widehat\bff^*_{\bf N}(-\bfk),\\
\cF_d(a\bff+b\bfg)&=& a\widehat\bff+b\widehat\bfg,\\
\label{4.8}
\cF_d(\nabla \bff)&=&i\bfk\widehat\bff,\\
\label{4.9}
\cF_d^{-1}\cF_d(\bff)&=&\bff,\\
\label{4.10}
\cF_d(\bff\odot\bfg)_{\bf N}(\bfk)&=&\widehat\bff_{\bf N}(\bfk)\cdot\widehat\bfg_{\bf N}(\bfk),
\EEEQ
where
\BBEQ
\label{4.11}
(\bff\odot\bfg)(\bfx^{\bf n}_{\bf N}):=\sum_{\bf j}\bff(\bfx^{\bf n-\bf j}_{\bf N})\cdot\bfg(\bfx^{\bf j}_{\bf N}).
\EEEQ
The operation described is a circular (or periodic) convolution. One important property of circular convolution is that it is commutative, meaning the order of the sequences does not affect the result:
\BBEQ
\label{4.12}
(\bff\odot\bfg)(\bfx^{\bf n}_{\bf N})=(\bfg\odot\bff)(\bfx^{\bf n}_{\bf N}).
\EEEQ
Furthermore, scalar multiplication distributes over convolution, so a constant factor can be applied to either operand without changing the outcome:
\BBEQ
\label{4.13}
a(\bff\odot\bfg)(\bfx^{\bf n}_{\bf N})=((a\bff)\odot\bfg)(\bfx^{\bf n}_{\bf N})=(\bff\odot(a\bfg))(\bfx^{\bf n}_{\bf N}).
\EEEQ
Among the most fundamental properties of the DFT is the convolution theorem, which transforms the convolution operation in physical (real) space into a simple pointwise multiplication in the Fourier domain—greatly simplifying numerical computation.
In the one-dimensional case, the circular convolution of two $N$-periodic sequences $f$ and $g$ results in a new $N$-length sequence $s$ defined as:
\BBEQ
\label{4.14}
s(n)=f(n)\odot g(n):=\sum_{m=0}^{N-1}f(m)g((n-m)\ {\rm mod}\ N),
%\sum_{l=0}^{N-1}f(l)g(n-l),
\EEEQ
where the modulo operation ensures periodicity by wrapping indices around the finite domain:
if $m = m_0 + lN$ with $m_0 \in {0, 1, \ldots, N-1}$ and $l \in \mathbb{Z}$, then $m \bmod N = m_0$.

Since the DFT input and output [Eqs. (\ref{4.3}) and (\ref{4.4}), respectively] are arrays of $ N=|\bfN |$ elements (real or complex numbers), they are similar to $N$--element vectors, and the DFT can be represented as the product of such a vector with an $N\times N$ matrix. Thus, the DFT [Eq. (\ref{4.3})] is a linear transformation of the discrete function $\bff_{\bf N}$ and can be written in matrix form as
\BBEQ
\label{4.15}
\widehat \bff_{\bf N}&=&\bfF\cdot\bff_{\bf N},
\EEEQ
where $\bfF$ is a square Vandermonde matrix constructed from $\omega^{-1}_{\bf N}$ [Eq. (\ref{4.5})], fully dense and invertible, with
\BBEQ
\label{4.16}
\bfF^{-1}=\frac{1}{|\bfN|}\bff(-\omega^{-1}).\nonumber
\EEEQ
In general, the DFT is a computationally costly algorithm, requiring $(O(N^2))$ operations, as each of the $N$ Fourier components depends on NNN input values. However, this cost can be drastically reduced to $O(N{\rm log}_2N)$ by exploiting the symmetries of the transformation matrix $\bfF$ [Eq. (\ref{4.15})], as achieved in the so-called Fast Fourier Transform (FFT) algorithm, first proposed by  \citep{Cooley`T`65} and further developed in subsequent works (see, e.g., 	\citep{Segurado`et`2018}). It should be noted that without the FFT, the DFT would be impractical for numerical methods; it is therefore the FFT, rather than the DFT itself, that gives its name to the numerical approaches briefly outlined in this section.

\section{FFT metods for CMs subjected to BFCS loading}

\setcounter{equation}{0}
\renewcommand{\theequation}{5.\arabic{equation}}

The original scheme by   \citep{Moulinec`S`1994} of FFT method, based on fixed-point (Picard) iterations, can, in a retrospective sense, be considered like an ignition spark for
a wide range of FFT-based methods for CMs.
So, the implicit Lippmann–Schwinger (L–S) equation (\ref{5.1}) (see also Eq. (\ref{2.24})) arising in linea elastic micromechanics can be transformed into a simple multiplication operation in the Fourier space, as shown in Eq. (\ref{5.2})
\BBEQ
\label{5.1}
\bfep(\bfx)&=&\bfep^{w_\Gamma}+\!{\bfU}^{(0)}*\bftau(\bfx), \\
\label{5.2}
\widehat\bfep(\bfk)&=&\!\widehat{\bfU}^{(0)}(\bfk)
\widehat\bftau(\bfk) \ \  (\bfk\not={\bf 0}),\ \ \  \widehat{\bfep}({\bf 0})={\bfep}^{w_\Gamma}.
\EEEQ

Equation (\ref{5.1})
is formulated under the periodic boundary conditions (PBC) given by (\ref{2.22}) at UC $\Omega_{00}$. 
If the PBC are replaced with the BFCS loading (\ref{2.5}) with the corresponding at PBC (\ref{2.19}) at BFUC $\Omega_{00}^b$, the classical formulation given by Eqs. (\ref{5.1}) and (\ref{5.2}) transforms accordingly into a new pair of modified L-S equations that reflect this loading framework
\BBEQ
\label{5.3}
\!\!\!\!\!\bfep(\bfx)\!\!&=&\!\!\bfep^{b(0)}(\bfx) +{\bfU}^{(0)}*\bftau(\bfx), \\
\label{5.4}
\!\!\!\widehat\bfep(\bfk)\!\!&=&\!\!\widehat{\bfep^{b(0)}}(\bfk)+\widehat\bfU^{(0)}(\bfk)
\widehat\bftau(\bfk), \ (\bfk\not={\bf 0}), 
\EEEQ
and $\widehat{\bfep}({\bf 0})=\widehat{\bfep^{b(0)}}({\bf 0})$. Under BFCS loading (\ref{2.5}), the field periodicity within the unit cell $\Omega_{00}^b$, which holds in Eq. (\ref{5.1}), is lost. As a result, 
$\Omega_{00}^b$ is replaced by a larger mesocell $\Omega_{00}^b$ that fully contains the representative volume element (RVE), i.e., ${\rm RVE}\subset \Omega_{00}^b$ (see Fig. 3).  
{\color{black} In other words, Eqs. (\ref{5.1}) and (\ref{5.3}) are formulated over different domains, $\Omega_{00}$ and $\Omega_{00}^b$, respectively, with $\Omega_{00}$ being a subset of $\Omega_{00}^b$ (i.e., $\Omega_{00} \subset \Omega_{00}^b$)..}
The size of the periodically distributed mesocells $\Omega_{00}^b$ acts as a postprocessing learning parameter, chosen so that the strain field $\bfep(\bfy)$ vanishes in the boundary layer region $\bfy \in \overline{\rm RVE}:=\Omega_{00}^b\setminus \text{RVE}$. This outer region imposes vanishing periodic boundary conditions, PBC (\ref{2.22}). 
Specifically, the vanishing of the strain field $\bfep(\bfy)$ in $\bfy\in \Omega_{00}^b\setminus $RVE enables $\bfep(\bfx)$ 
($\bfx\in \Omega_{00}^b$) to be regarded as periodic within an extended medium where the mesocell $\Omega_{00}^b$ serves as the new periodicity cell.
The resulting Eqs. (\ref{5.3}) and (\ref{5.4}) can be solved iteratively using the same Picard (fixed-point) method as used for (\ref{5.1}) and (\ref{5.2}). This approach yields estimates of the dataset $\bfcD^{\rm p}_k$ (\ref{3.7}) for composite materials (CMs) with deterministic structure--a class for which FFT methods have not previously been applied.
In this context, each specific translation $\bfchi\in {\cal V}_{\rm \bf x}$ (\ref{3.5}) is treated as a deterministic structure. Consequently, the dataset $\bfcD^{\rm p}_k$ (\ref{3.7}) is obtained by averaging over these translations, using the statistical formulation given in (\ref{3.5}) and (\ref{3.6}).

Just as the Picard iteration scheme for solving Eqs. (\ref{5.1}) and (\ref{5.2}) under periodic boundary conditions (PBC) (\ref{2.22}) formed the basis for FFT-based homogenization methods for periodic composites  \citep{{Buryachenko`2023j},  {Lucarini`et`2022},  {Schneider`2021},  {Segurado`et`2018}}, the analogous scheme for Eqs. (\ref{5.3}) and (\ref{5.4}) under BFCS loading (\ref{2.5}) may similarly drive the development of a new generation of FFT approaches. These would extend existing methods to handle both deterministic and periodic structures within a unified framework, accommodating non-periodic loading in $\Omega_{00}$ while preserving the efficiency of Fourier-based solvers for $\Omega_{00}^b$. Deterministic microstructures could be embedded in mesocells with BFCS loading, enabling FFT analysis beyond classical PBC assumptions. For periodic structures, statistical averaging over translations (via dataset $\bfcD^{\rm p}_k$ (\ref{3.7})) naturally integrates with this scheme. As with the original FFT method by \citep{Moulinec`S`1994}, the proposed extension promises significant advances in modeling composites with high contrast, nonlinearity, or nonlocal effects.

So, by the use of the DFT's properties (\ref{4.9}) and (\ref{4.10}), the discretized modified L-S equation (\ref{5.3}) can be presented in the
form (see  \citep{Zeman`et`2010} for similar manipulation of L-S equation (\ref{5.1}) 
\BBEQ
\label{5.5}
(\bfcI-\bfcB^{U})\bfe=\bfe^{b(0)}, \ \ \ \bfcB^{U}={\cal F}_d^{-1}\widehat{\bfcU}{\cal F}_d\bfcL_1
\EEEQ
where all the matrices exhibit a block-diagonal structure, e.g., $\widehat{\bfcU}=[\delta_{\bf n m}\widehat\bfU^{(0){\bf n m}}_{\alpha \beta}],$
$\bfcI=[\delta_{\bf n m}\delta_{\alpha \beta}]$, ${\bfcL_1}=[\delta_{\bf n m}\bfL^{\bf n m}_{1|\alpha \beta}],$ ,
($\bfn,\bfm\in Z^d_{\bf N}$, $\alpha,\beta=1,\ldots,d$) whereas $\bfe=(\bfep^{\bf n}_{\alpha})$ and
$\bfe^{b(0)}=((\bfep^{b(0)})^{\bf n}_{\alpha})$ ($\bfn\in Z^d_{\bf N}$, $\alpha=1,\ldots,d$). 
The cost of multiplication by $\bfcB^{U}$ is governed by the forward DFT and inverse iDFT, both efficiently performed in $O(|\bfN|\log|\bfN|)$ operations using the FFT. This makes system (\ref{5.5}) well suited to iterative schemes. In particular, the original FFT-based basic scheme proposed in  \citep{Moulinec`S`1994} expresses the solution of (\ref{5.5}) as a Neumann series of the matrix inverse 
$(\bfcI-\bfcB^U)^{-1}$:
\BBEQ
\label{5.6}
\bfe^{[j]}=\sum_{p=0}^j(\bfcB^{U})^{p}\bfe^{b(0)}.
\EEEQ
which is similar in form to the fixed-point iterative method
(Picard’s iterations) ($\bfep^{[0]}(\bfx)=\bfep^{b(0)}$)
%\noindent 

\vspace{.5mm}
\hspace{2.0cm} {\bf Algorithm 1.} {\it Modified basic scheme}
\vspace{-1.3cm}

\BB
\hspace{-5.cm} \begin{picture}(100, 50)
\put(20,20){\line(1,0){250}} % horizontal line
\end{picture}
\nonumber 
\EE
\vspace{-1.89cm}

\begin{align}
&{\bf Data:} \bfL^{(0)}, \bfL(\bfx), {\rm tol},\bfep^{b(0)}(\bfx) \nonumber \\[-2pt]
& {\bf Result:} \ \bfep(\bfx) \nonumber \\[-2pt]
&\bfep^{[0]}= \bfep^{b(0)}(\bfx) \nonumber \\[-2pt]
&{\bf while:}\ \Delta_{\sigma}>{\rm tol} \ \ {\bf do} \nonumber \\[-2pt]
\label{5.7}
& \hspace{0.8cm} \bftau^{[j]}(\bfx) =\!\bfL_1(\bfx)\bfep^{[j]}(\bfx) \\[-2pt]
\label{5.8}
& \hspace{0.8cm} \widehat\bftau^{[j]}(\bfk)={\cal F}_d({\bftau^{[j]}}(\bfx)),\\[-2pt]
\label{5.9}
& \hspace{0.8cm} \widehat\bfep^{[j+1]}(\bfk)= {\widehat\bfU}^{(0)}(\bfk){^L\!\widehat\bftau}^{[j]}(\bfk),\\[-2pt]
\label{5.10}
& \hspace{0.8cm} \bfep^{[j+1]}(\bfx)=\bfcF_d^{-1}\big({\widehat\bfep}^{[j+1]}(\bfk)\big) +\bfep^{b(0)}, \\[-2pt]
& {\rm end}\nonumber 
\end{align}

\vspace{-1.6cm}
\BB
\hspace{-5.cm} \begin{picture}(100, 50)
\put(20,20){\line(1,0){250}} % horizontal line
\end{picture}
\nonumber 
\EE

\vspace{-0.6cm}

\noindent where $\bfk\not={\bf 0}$ and $\widehat\bfep^{[j+1]}({\bf 0})=\widehat{\bfep}^{b(0)}({\bf 0})$ in Eq. (\ref{5.9}). The convergence behavior of the modified basic scheme applied to Eqs. (\ref{5.3}),(\ref{5.4}) is the same as the original basic scheme used for Eqs. (\ref{5.1}),(\ref{5.2}). A common stopping criterion, first introduced by \citep{Moulinec`S`1998}, is based on the relative equilibrium residual:
\BBEQ
\label{5.11}
\Delta_{\sigma}:=\frac{|\nabla\cdot\bfsi|_{L_2}}{|\Omega_{00}^b\lle\bfsi\rle|}=\frac{\big[\sum_{\xi}|\bfxi\cdot\bfsi|^2\big]^{1/2}}
{|\widehat\bfsi({\bf 0})|}<{\rm tol},
\EEEQ
where $||\bullet||_{L_2}$ represent the $L_2$-norm of the vector field and $||\bullet||$ is the Frobenius norm of the tensor.
By Parseval’s theorem, this criterion is efficiently evaluated in Fourier space, making it well-suited for FFT-based iterative solvers.

Thus, the modified L-S equation (\ref{5.3}) can be solved in Fourier space similarly to a pointwise product, by using the convolution property (\ref{4.10}), which reduces it to Eq. (\ref{5.9}). In this form, the polarization $\bftau$ (\ref{5.9}) depends on the strain field $\bfep(\bfx)$ and local stiffness $\bfL(\bfx)$ in real space $\bfx\in R^d$, requiring both forward (\ref{5.8}) and inverse (\ref{5.10}) DFTs. \citep{Brisard`D`2010} (see also  \citep{Schneider`2021}) interpreted the discretization by \citep{Moulinec`S`1994}
 as similar to an under-integrated conforming Galerkin discretization of the Hashin–Shtrikman variational principle, with voxel-wise constant strain ansatz functions.

An alternative approach to enhance the convergence of FFT-based solvers, beyond the basic scheme, was introduced by 
\citep{Zeman`et`2010} and \citep{Vondrejc`et`2012} in the context of vector field homogenization for electrostatics. In this formulation, the integral L-S equation (\ref{5.1}) is discretized using the trigonometric collocation method, wherein trigonometric polynomials serve as basis functions for interpolating field quantities in real space. The projection of the continuous equation onto this discrete function space yields a linear system of equations, where the nodal values of the strain field are treated as the primary unknowns. This discrete system is structurally similar to that arising in Galerkin-type methods and can be efficiently solved using Krylov subspace algorithms, such as the conjugate gradient (CG) or biconjugate gradient (BiCG) methods by \citep{Zeman`et`2010}. The resulting discrete counterpart of the modified L-M equation (\ref{5.3}) takes the form:
\BBEQ
\label{5.12}
\bfep(\bfx)-{\cal F}_d^{-1}\Big\{\bfU^{(0)}(\bfxi):{\cal F}_d\big[\bfL_1(\bfx):\bfep(\bfx)\big]\Big\}=\bfep^{b(0)}(\bfx),\
\EEEQ
Here, $\bfep(\bfx)$ denotes the discrete (nodal) representation of the strain field. The left-hand side of Eq. (\ref{5.12}) defines a linear operator acting on $\bfep$, which can be directly utilized within a Krylov subspace solver. The structure of this linear operator is similar to that required by iterative methods such as the conjugate gradient algorithm. The complete procedure for the Krylov-based scheme is summarized in Algorithm 2.

\vspace{1.5mm}
\hspace{2.0cm} {\bf Algorithm 2.} {\it Modified Krylov-based scheme}
\vspace{-1.3cm}

\BB
\hspace{-5.cm} \begin{picture}(100, 50)
\put(20,20){\line(1,0){250}} % horizontal line
\end{picture}
\nonumber 
\EE
%\vspace{-3.2cm}
\vspace{-1.89cm}

\begin{align}
&{\bf Data:} \bfL^{(0)}, \bfL(\bfx), {\rm tol},\bfep^{b(0)}(\bfx) \nonumber \\[-2pt]
& {\bf Result:} \ \bfep(\bfx) \nonumber \\[-2pt]
&\bfep={\bf 0} \nonumber \\[-2pt]
&{\rm Solve}\ \bfcA(\bfep(\bfx))=\bfep^{b(0)}(\bfx)\ {\rm by\ conjugate
\ gradients\ with\ tol} 
\nonumber \\[-2pt]
\label{5.13}
& \hspace{0.8cm} \bfcA(\bfep(\bfx))= \bfep(\bfx)-{\cal F}_d^{-1}\Big\{\bfU^{(0)}(\bfxi):{\cal F}_d\big[\bfL_1(\bfx):\bfep(\bfx)\big]\Big\} 
%& {\rm end}\nonumber 
\end{align}

\vspace{-1.6cm}
\BB
\hspace{-5.cm} \begin{picture}(100, 50)
\put(20,20){\line(1,0){250}} % horizontal line
\end{picture}
\nonumber 
\EE
\vspace{-1.2cm}

Unlike traditional strain-based FFT schemes, displacement-based approaches solve directly for the displacement field, reducing memory usage and yielding a determinate system. This avoids the rank-deficiency issues of strain-based formulations and enables the use of preconditioners and alternative iterative solvers.

An alternative   {\it displacement-based} FFT formulation for the modified L-S equation (\ref{5.3}--extending a similar approach developed for Eq. (\ref{5.1}) (see  \citep{{Lucarini`S`2019},  {Lucarini`et`2022},  {Schneider`2021}})--focuses on
directly solving for the displacement field instead of the strain. By substituting the displacement decomposition (\ref{2.10}) into the constitutive relation (\ref{2.2}), the equilibrium equation (\ref{2.1}) can be reformulated accordingly
\BBEQ
\label{5.14}
\nabla\cdot[\bfL(\bfx):\nabla^s\bfu_1(\bfx)]= -\nabla\cdot[\bfL(\bfx):\bfep^{b(0)}].
\EEEQ
This approach eliminates the need for a reference medium and supports both standard and staggered discretizations via discrete differential operators. The key idea is to derive a fully determined system in Fourier space by expressing the problem in terms of displacement fluctuations at each grid point. By removing symmetries associated with the real Fourier transform and excluding the zero-frequency component, the formulation yields a determinate linear system suitable for preconditioning.
Rather than introducing a reference medium—as in Eq. (\ref{2.13})—to express the microstructural dependence $\bfL(\bfx)$ through an eigenstrain, Eq. (\ref{5.14}) is directly transformed into Fourier space to compute spatial derivatives. Thus, applying the Fourier transform to the linear momentum conservation equation (\ref{5.14}) yields
\BBEQ
\label{5.15}
\widehat{\bfbd}:\cF_d\big[^L\!\bfL(\bfx):\cF^{-1}_d\big(\widehat{\bfbs}\cdot\widehat\bfu_1\big)\big] =-\widehat{\bfbd}:\cF_d\big[\bfL(\bfx):\bfep^{b(0)}\big],
\EEEQ
Using the Fourier derivative property (\ref{4.8}), the operators in Fourier space are defined as
\BBEQ
\label{5.16}
\widehat{\bfbd}(\bfk):=\widehat d_{pqr}(\bfk)=ik_r\delta_{pq},\ \ \
\widehat{\bfbs}(\bfk):=\widehat s_{pqr}(\bfk)=\frac{1}{2}(ik_q\delta_{pr}+ik_p\delta_{qr}),
\EEEQ
which correspond to the divergence and symmetric gradient operators, respectively, with dependence on the frequency vector 
$\bfk$. Thus, both the differential form of the equilibrium equation (\ref{5.15}) and its integral counterpart (\ref{5.3}) reduce, under the DFT, to algebraic operations in Fourier space. When the Fourier transform is discretized as a DFT, Eq. (\ref{5.16}) becomes a linear system over complex variables with $\bfu_1$ 
as the primary unknown in Fourier space. This system can be efficiently solved using either direct or iterative solvers, potentially with preconditioning in Fourier space (see  \citep{Lucarini`S`2019}).

It is important to emphasize that the linear system formulated in Eq. (\ref{5.15}) is not expressed in terms of a traditional matrix of coefficients, but rather through the action of a linear operator. This system can be formally written as
\BBEQ
\label{5.17}
\bfcA(\bfu_1) = \bfb^{\cal A},
\EEEQ
where $\bfcA$ denotes a linear operator acting on the vector $\bfu_1$, and $\bfb^{\cal A}$ represents the known right-hand side derived from Eq. (\ref{5.15}). Given the operator-based structure of Eqs. (\ref{5.17}), classical matrix-based solvers are not directly applicable. Among the various iterative methods suitable for operator-defined systems, the Conjugate Gradient (CG) method is particularly effective and is employed in this study. As discussed by  \citep{Barrett`et`1994} (see also 
 \citep{Brisard`D`2010}), the key requirement for implementing the CG method in this context is the ability to compute the action of the operator $\bfcA$ on a given input vector $\bfu_1$, without explicitly assembling a global stiffness matrix. This matrix-free formulation is highly advantageous in terms of memory efficiency and scalability. The evaluation of the operator action $\bfcA(\bfu_1)$ involves a structured sequence of operations that leverage the FFT framework. Naturally, the FFT is efficiently employed to perform the required discrete Fourier transforms (DFTs) and their inverses (iDFTs). 
For completeness, we provide below the pseudocode for the standard (unpreconditioned) Conjugate Gradient (CG) algorithm:
%\vspace{-0.3cm}
The method advances by iteratively constructing sequences of vectors: the approximations to the displacement fluctuation field
$\bfu_1^{[j]}$ (line (\ref{5.19}) in Algorithm 3), the corresponding residuals $\bfr^{[j]}$ (line (line (\ref{5.19})), and the conjugate (search) directions $\bfp^{[j]}$ used to update both iterates and residuals. Each iteration involves two key scalar products (lines (\ref{5.18}) and (\ref{5.21})) that yield step size parameters ensuring the satisfaction of orthogonality and conjugacy conditions between residuals and search directions. It is worth noting that the primary computational expense per iteration of the CG algorithm stems from the evaluation of the operator $\bfcA(\bfu_1)$. This operation entails the same number of forward and inverse FFTs as a single iteration of the modified basic scheme.. Consequently, the computational cost of applying $\bfcA$ is largely governed by the discrete Fourier transform operations $\cF_d$ and $\cF_d^{-1}$, both of which scale as $O(|\bfN|\log |\bfN|)$ due to the efficiency of FFT algorithms.

\vspace{1.5mm}
\hspace{2.0cm} {\bf Algorithm 3.} {\it Modified CD scheme}
\vspace{-1.3cm}

\BB
\hspace{-5.cm} \begin{picture}(100, 50)
\put(20,20){\line(1,0){250}} % horizontal line
\end{picture}
\nonumber 
\EE
%\vspace{-3.3cm}
\vspace{-1.89cm}

\begin{align}
&{\bf Data:}\ \ \bfL(\bfx), {\rm tol},\bfep^{b(0)}(\bfx) \nonumber \\[-2pt]
& {\bf Result:} \ \bfep(\bfx) \nonumber \\[-2pt]
&\bfu_1^{[0]}= \bfb^{\cal A},\ \ \bfr^{[0]}=\bfb^{\cal A}-\bfcA(\bfu_1^{[0]}, \ \ \bfp^{[0]}=\bfr^{[0]} \nonumber \\[-2pt]
& {\bf for} \ \ j=0,1,2,\ldots, \ {\rm until \ convergence} \ \ {\bf do} \nonumber \\[-2pt]
\label{5.18}
& \hspace{0.8cm} \alpha^{[j]}:=(\bfr^{[j]\top}\bfr^{[j]})/(\bfp^{[j]\top}\bfcA^{\top}(\bfp^{[j]})) \\[-2pt]
\label{5.19}
& \hspace{0.8cm} \bfu_1^{[j+1]}:=\bfu_1^{[j]}+\alpha^{[j]}\bfp^{[j]} \\[-2pt]
\label{5.20}
& \hspace{0.8cm} \bfr^{[j+1]}:=\bfr^{[j]}-\alpha^{[j]}\bfcA(\bfp^{[j]}) \\[-2pt]
\label{5.21}
& \hspace{0.8cm} \beta^{[j]}:=(\bfr^{[j+1]\top}\bfr^{[j+1]})/(\bfr^{[j]\top}\bfr^{[j]}) \\[-2pt]
\label{5.22}
& \hspace{0.8cm} \bfp^{[j+1]}:=\bfr^{[j+1]}+\beta^{[j]}\bfp^{[j]} \\[-2pt]
& {\bf end\ for} \nonumber \\[-2pt]
&\bfep(\bfx)=\bfu_1(\bfx)+\bfep^{b(0)}(\bfx). \nonumber
\end{align}

\vspace{-1.5cm}
\BB
\hspace{-5.cm} \begin{picture}(100, 50)
\put(20,20){\line(1,0){250}} % horizontal line
\end{picture}
\nonumber 
\EE
\vspace{-1.0cm}

In the classical setting of symmetric positive-definite (SPD) systems, these conditions ensure that each new iterate minimizes the error in an energy norm, and convergence is achieved in at most $n$ steps for an $n$-dimensional problem.
Interestingly, \citep{Vondrejc`et`2012} have shown that the CG algorithm remains effective within the local micromechanical framework even when the system matrix—such as in Eq. (\ref{5.5})—is not strictly symmetric. This empirical robustness opens the door for broader applications of CG in FFT-based schemes, including those where standard SPD assumptions do not hold. This insight is particularly relevant in the context of micromechanical formulations where the governing operators often arise from variational or weak forms that yield non-symmetric but still well-conditioned systems.
Furthermore, within each CG iteration, a few inner iterations are performed to solve the modified L-S formulation (see Eqs. (\ref{5.3}) and (\ref{5.4})). These steps closely follow the structure of the classical basic scheme of the L-S equation (\ref{5.1}) and are embedded within the Krylov solver framework, similar to the nested iterative strategy outlined by \citep{Lahellec`et`2003}.
In the context of the L-S equation (\ref{5.1}) (and also of the modified L-S equation (\ref{5.3})), numerical experiments confirm the robustness of this approach with respect to variations in internal parameters such as mesh resolution and material heterogeneity. Notably, the method demonstrates markedly improved convergence rates for problems characterized by high-contrast material coefficients—while maintaining a low per-iteration computational overhead due to the efficient use of FFT-based operator evaluations.

This establishes a formal correspondence between the discrete Fourier transforms (DFT) of the classical L-S equation (\ref{5.2}) and its modified counterpart (\ref{5.4}), along with their respective equilibrium formulations (cf.  \citep{Lucarini`et`2022} and Algorithm 3). This structural similarity enables the extension of numerous FFT-based algorithms—originally developed over the past three decades for micromechanics of periodic media under PBC (\ref{2.22}), starting from the pioneering work by \cite{Moulinec`S`1994} (see also the comprehensive reviews  \citep{{Lucarini`et`2022},  {Schneider`2021}})—to analogous formulations that accommodate BFCS loading (\ref{2.5}) under the different PBC (\ref{2.19}). 
Since the fields vanish in the surrounding buffer zone $\overline{\rm RVE} := \Omega_{00}^b \setminus {\rm RVE}$, i.e., $|\bfsi(\bfy)|, |\bfep(\bfy)| < {\rm tol}$ for $\bfy \in \overline{\rm RVE}$, the internal field distributions $\bfsi(\bfx)$ and $\bfep(\bfx)$ within the domain $\bfx \in {\rm RVE}$ remain unaffected by further enlargement of the body-force unit cell (BFUC) $\Omega_{00}^b$. Consequently, the FFT-based solution of the modified L-S equation (\ref{5.3}) is identical for composite materials with either periodic ($\Omega_{00} \ne \mathbb{R}^d$) or deterministic ($\Omega_{00} = \mathbb{R}^d$) microstructures.
In contrast, composite images obtained from micro-computed tomography (micro-CT) or scanning electron microscopy (SEM) typically contain several hundred or even thousands of inclusions, see  \cite{{Bellens`et`2024},{Echlin`et`2014}}. Importantly, such CT-derived images can be interpreted as observational snapshots (or ``windows of observation") of deterministic composite structures. The observation window $w$ is subjected to boundary conditions (either (\ref{2.20}$_1$), (\ref{2.20}$_2$), or (\ref{2.22})). To enable numerical simulation using FEM or FFT methods, periodization of $w$ is usually performed by rearranging inclusions near the boundary $\partial w$, which inevitably introduces boundary layer effects. In contrast, within the new proposed RVE framework, a similar rearrangement is applied only in regions $\overline{\rm RVE}$ where the fields vanish. As a result, boundary layer effects are inherently and permanently eliminated.

Furthermore,  upon projection of the equilibrium equations to a discrete functional space, the governing equations reduce to a linear system in which the unknown is the discrete, nodal representation of the displacement fluctuation field $\bfu_1$. This system can be efficiently solved using iterative Krylov subspace methods, such as the Conjugate Gradient (CG) or Biconjugate Gradient (BiCG) algorithms, which are well-suited to large-scale, sparse systems. Specifically, the left-hand side of Eq. (\ref{5.17}), $\bfcA(\bfu_1)$, represents the application of a linear operator (rather than an explicit matrix) to the field $\bfu_1(\bfx)$ and thus lends itself naturally to matrix-free implementations. This operator can be directly utilized in Krylov solvers, including those provided by high-performance computing libraries such as PETSc  \citep{Balay`et`2016} (see also  \citep{Plimpton`et`2018}). 
Importantly, in the context of the L-S equation (\ref{5.1}) (the modified L-S equation (\ref{5.3}) can be similarly considered), the studies  by \citep{Zeman`et`2010} and  \citep{Vondrejc`et`2012} have demonstrated that, in contrast to traditional acceleration techniques for FFT-based schemes (e.g.,  \citep{Eyre`M`1999} and  \citep{Michel`et`2001} methods), this Krylov-based method maintains the same per-iteration computational complexity as the original basic scheme.
. That is, each iteration incurs a cost comparable to that of a single application of the modified L-S operator in the basic FFT scheme, primarily driven by the required FFT and inverse FFT operations. Furthermore, the convergence behavior of the Krylov-based approach is shown to be largely independent of the choice of the auxiliary reference medium $\bfL^{(0)}$, eliminating the need for fine-tuning this parameter. The method offers significantly improved convergence rates over the modified basic scheme (\ref{5.7})–(\ref{5.9}), particularly for moderate stiffness contrasts. However, its performance still degrades in scenarios involving high contrast in material stiffness. In such cases, the condition number of the system matrix associated with Eq. (\ref{5.3}) becomes excessively large, rendering the system ill-posed and causing the iterative solver to converge slowly or stagnate. Despite this, the algorithm remains robust and efficient for a broad class of heterogeneous media and stands out for its simplicity, generality, and ability to leverage existing high-performance Krylov solvers.

{\color{black} In broad terms, FFT methods can be grouped into two categories. The first category originates from the L–S equations (\ref{5.1}) and (\ref{5.2}), while the second is derived from the ML–S equations (\ref{5.3}) and (\ref{5.4}). The most apparent distinction lies in the respective representative computational domains, UC $\Omega_{00}$ and BFUC $\Omega_{00}^b$. A more essential difference concerns the intended outcomes of solving Eqs. (\ref{5.1}), (\ref{5.2}) versus Eqs. (\ref{5.3}), (\ref{5.4}). Solving (\ref{5.1}) and (\ref{5.2}) provides a complete solution, yielding both DNS results and the effective moduli $\bfL^*$ (\ref{6.3}). By contrast, solving (\ref{5.3}) and (\ref{5.4}) also serves the additional purpose of determining the RVE. If the RVE is contained within $\Omega_{00}^b$, the resulting DNS is then employed to construct $\bfcD^{\rm p}$.
If the RVE is not fully contained within RVE$\not\subset\Omega_{00}^b$, its size $\Omega_{00}^b$ must be increased, and the solution of Eqs. (\ref{5.3}) and (\ref{5.4}) repeated. It is also important to clearly distinguish between the BFCS $\bfb(\bfx)$ in (\ref{2.5}) and the periodic $\bfb(\bfx)$ in (\ref{2.7}). The body force $\bfb(\bfx)$ has compact support $\bfcB^b$ when the identified RVE lies within $\Omega_{00}^b$. In this case, the solution $\bfep(\bfx)$ in the isolated domain $\Omega_{00}^b$ (associated with the BFCS $\bfb(\bfx)$ from (\ref{2.5})) is identical to the solution in the periodic medium with the representative BFUC $\Omega_{00}^b$ (associated with the periodic body force $\bfb(\bfx)$ from (\ref{2.7})). 
Thus, if and only if RVE$\subset \Omega_{00}^b$ (since the condition $\bfcB^b\subset \Omega_{00}^b$ is not sufficient), the notions of the  BFCS $\bfb(\bfx)$   (\ref{2.5}) and periodic body force $\bfb(\bfx)$ (\ref{2.7}) 
coincide.  
This equivalence enables the generalization of the FFM methods to solve the ML–S equation 
 (\ref{2.14}) -- rather than L-S equation (\ref{2.24}), see \citep{{Brisard`D`2010},{deGeus`et`2017},{LiM`et`2024},{Lucarini`S`2019},{Michel`et`2001},{Moulinec`S`1994},  {Moulinec`S`1998}, {Segurado`et`2018},{Schneider`2021},{Wang`et`2024},{Wicht`et`2021},{Lucarini`et`2022},{Zeman`et`2010},{Zeman`et`2017}}.
}

\section{Effective elastic moduli and surrogate operators }
\setcounter{equation}{0}
\renewcommand{\theequation}{6.\arabic{equation}}

Let us consider periodic CM with inclousion centers $\Lambda_{00}$ and PBC (\ref{2.22}) at UC $\Omega_{00}$.
. By the use of the Gauss-Ostrogradsky theorem, we define the overall macrostress $\{\bfsi\}=\lle\bfsi\rle^{\Omega}$, and the overall macrostrain $\{\bfep\}=\lle\bfep\rle^{\Omega}$ of the UC $\Omega_{00}$
\BBEQ
\label{6.1}
\lle\bfsi\rle^{\Omega}\!\!\!&:=&\!\!\! |{\Omega_{00}}|^{-1}\int_{\Omega_{00}}\bfsi(\bfx)d\bfx=
|{\Omega_{00}}|^{-1}\int_{\Gamma^0}\bft(\bfs)\,\,\,^{^S}\!\!\!\!\!\! \otimes\bfs d\bfs,\\
\label{6.2}
\lle\bfep\rle^{\Omega}\!\!\!&:=&\!\!\! |{\Omega_{00}}|^{-1}\int_{\Omega_{00}}\bfep(\bfx)d\bfx=
|{\Omega_{00}}|^{-1}\int_{\Gamma^0}\bfu(\bfs)\,\,\,^{^S}\!\!\!\!\!\! \otimes\bfn(\bfs) d\bfs,
\EEEQ
in terms of the traction $\bft(\bfs):=\bfsi(\bfs)\bfn(\bfs)$ and the displacement $\bfu(\bfs)$ on the geometrical boundary of the UC $\bfs\in \Gamma^0$ with the outward normal unit vectors $\bfn(\bfs)$ on $\Gamma^0$.
The effective stiffness $\bfL^*$ is estimated as a proportionality factor
between the UC's averages of the stresses $\lle\bfsi\rle^{\Omega}$ and strains $\lle\bfep\rle^{\Omega}$
\BB
\label{6.3}
\lle\bfsi\rle^{\Omega}=\bfL^*\lle\bfep\rle^{\Omega}.
\EE
Estimating the involved macro variables is performed by the {{\it micro-to-macro} transition.
Evaluations of effective moduli (\ref{6.3}) by the FFT methods are well developed directions in micromechanics (see Introduction for references). 

We now focus on body force with compact support (BFCS), as described by Eq. (\ref{2.5}). The foundational work of Silling and co-authors  \citep{{Silling`2020},  {You`et`2020},  {You`et`2024}} introduced machine learning (ML) approaches to develop surrogate nonlocal operators for CMs, using datasets generated via DNS. Specifically, their dataset structure is given by
\BBEQ
\label{6.4}
\!\!\!\!\!\!\!\!\!\!\!{\bfcD}^{\rm DNS}=\{\bfcD^{\rm DNS}_k\}_{k=1}^N, \ 
\bfcD^{\rm DNS}_k=\{\bfu(\bfb_k,\bfx),\bfb_k(\bfx)\},
\EEEQ
where each realization corresponds to a different body force loading $\bfb_k(\bfx)$. Their work primarily focused on 1D heterogeneous bars under wave-like loading at the boundary and oscillatory body forces. In contrast, the current study (see  \citep{{Buryachenko`2023},  {Buryachenko`2023a}}) extends this idea to more general periodic and deterministic microstructures by
replacing the DNS dataset $\bfcD^{\rm DNS}$ (\ref{6.4}) with datasets 
$\bfcD^{\rm p}$ (\ref{3.7}) and $\bfcD^{\rm d}$ (\ref{3.8}), 
corresponding to periodic and deterministic configurations, respectively. Here, the BFCS $\bfb(\bfx)$ serves both as a means of loading and as an input to infer nonlocal constitutive behavior. The datasets are compressed into surrogate forms $\widetilde\bfcD^{\rm p}$ and $\widetilde\bfcD^{\rm d}$ to reduce computational cost while preserving essential micromechanical detail. 
Each of these surrogate datasets is used to learn a corresponding nonlocal operator $\bfcL_\gamma$, characterized by a convolution-type integral operator: 
\vspace{-2.mm}
\BBEQ
\!\!\!\!\!\!\!\!\!\!\!\!\!\!\bfcL_{\rm \gamma}[\lle{\bfu}_k\rle](\bfx) &=& {\bf \Gamma}(\bfx), \nonumber\\
\label{6.5}
\!\!\!\!\!\!\!\!\! \!\!\!\!\!\bfcL_{\rm \gamma}[\lle{\bfu}_k\rle](\bfx) &=&
\!\!\int \!\!\bfK_{\gamma}(|\bfx-\bfy|) (\lle{\bfu}_k\rle(\bfy)-\lle{\bfu}_k\rle(\bfx))~d\bfy,
\EEEQ
where ${\bf \gamma}=b,\sigma,\epsilon i,\sigma i$ indexes one of four different operator types—associated with either displacement, stress, or their localized forms—and ${\bf \Gamma}_k(\bfx)$ represents the corresponding averaged response: $-\bfb_k(\bfx)$, $\langle \bfsi \rangle(\bfx)$, $\langle \bfep \rangle_i(\bfx)$, or $\langle \bfsi\rangle_i(\bfx)$. To determine the optimal kernel $\bfK_\gamma^*$ for each case, the following minimization problem is solved:
\BBEQ
\label{6.6}
\!\!\!\!\!\!\!\!\!\!\!\!\!\!\!\!\!\!\!\!\!\bfK_{\gamma}^*={\rm arg}\!\min_{\!\!\!\!\!\!\!\!\!\! {{\bf K}_{\gamma}}}\!\sum_{k=1}^N\!|| \bfcL_{\rm {\gamma}}[\lle{\bfu_k}(\bfb_k)\rle](\bfx)- {\bf\Gamma}_k(\bfx)||^2_{l_2} %\nonumber\\
%\!\!&+&\!\!
+{\cal R}(\bfK_{\gamma}).
\EEEQ
where $\mathcal{R}(\bfK_\gamma)$ is a regularization term (e.g., Tikhonov regularization) to stabilize the inverse problem. The kernel $\bfK_\gamma$ is parameterized using Bernstein polynomial bases, and the optimization is performed using the Adam algorithm  \citep{Kingma`B`2014}. 
This formulation generalizes and strengthens earlier works  \citep{{You`et`2020},  {Fan`et`2023},  {You`et`2021},  {You`et`2022}, 
 {You`et`2024}}, providing a framework that combines physically motivated loading, rigorous micromechanical averaging, and data-driven learning of nonlocal operators across a wide class of complex materials.

The methods by \citep {You`et`2020} and  \citep{You`et`2024} rely on uncompressed DNS datasets (\ref{6.4}), which store full microscale displacement fields for each BFCS $\bfb_k(\bfx)$, resulting in large data volumes. In contrast, the compressed datasets ${\cal D}^{\rm I}$ ($I=p,d$) —whether periodic or deterministic—require no full-field DNS and instead use micromechanics-based averaging to extract effective quantities more efficiently. 
For a linearized homogeneous peridynamic medium under remote homogeneous BCs ((\ref{2.20}$_1$) or (\ref{2.20}$_2$)), classical peridynamics yields local moduli directly from the constitutive relation  \citep{Silling`et`2003}. Similarly, for a surrogate homogeneous medium under the same BCs, the effective stiffness tensor is given by
\BBEQ
\label{6.7}
\bfL^*&=&\int\bfK_{\sigma}(|\bfx-\bfy|)(\bfy-\bfx)~d\bfy,
\EEEQ
providing a compact expression for the homogenized elastic response. An analogy of the strain concentration (\ref{3.2}) can be expressed as
\BBEQ
\label{6.8}
\lle\bfep\rle_i(\bfx)=\int\bfK_{\epsilon i}(|\bfx-\bfy|)(\bfy-\bfx)~d\bfy\lle\bfep\rle.
\EEEQ

The surrogate operators in Eqs. (\ref{6.5}) and (\ref{6.6}) are fixed and limited to modeling linear responses. To overcome this limitation, nonlocal neural operators have been proposed to learn general mappings between function spaces, offering greater flexibility and adaptability  \citep{{Lanthaler`et`2024},  {Li`et`2003}}.
Traditional artificial neural networks (ANNs), such as fully connected neural networks (FCNNs), define local nonlinear operators. For example, an $L$-layer FCNN $\Psi(\bfx)$: $\bfR^{\rm d_{\bf x}}\to \bfR^{\rm d_{\rm \bf u}}$ maps input $\bfx$ to output $\bfu$ through successive transformations:
\BBEQ
\label{6.9}
%\nonumber
\!\!\!\!\!\!\!\!\!\!\!\!\!\!\!\!\!\bfz^l(\bfx)=\bfcA(\bfw^l \bfz^{l-1}(\bfx)+\bfb^l), \ \bfu(\bfx)=\bfw^L \bfz^{L-1}(\bfx)+\bfb^L,
\EEEQ
where $\bfcA$ is a nonlinear activation function (e.g., ReLU or tanh), and the learnable parameters are $\bfthe = \{\bfw^l, \bfb^l\}_{l=1}^L$. Importantly, this type of operator is local, since the output at a point $\bfx$ depends only on the input at that same point.
In contrast, nonlocal neural operators generalize this by incorporating integral terms to model spatial interactions and long-range dependencies. For instance, a typical nonlocal layer takes the form:
\BB
\label{6.10}
\bfz^l(\bfx)= \bfcA(\bfw^l \bfz^{l-1}(\bfx)+\bfb^l+(\bfcK^l(\bfz^{l-1})(\bfx)),
\EE
where $\bfcK^l$ represents an integral operator involving a learnable kernel $\bfK^l$, enabling the model to aggregate information from across the spatial domain. 
A wide range of architectures has been developed based on this idea, including Deep Operator Networks (DeepONet), PCA-Net, Graph Neural Operators, Fourier Neural Operators (FNO), and Laplace Neural Operators (LNO).
These methods differ in how they implement the nonlocal interactions and the structure of the kernel $\bfK^l$. Comparative reviews and benchmarks of these neural operator frameworks are available in  \citep{{Gosmani`et`2022},  {HuZ`et`2024},  {Kumara`Y`2023},  {Lanthaler`et`2024}}.

Physics-Informed Neural Networks (PINNs)  \citep{{Cuomo`et`2022},  {Haghighata`et`2021},  {Harandi`et`2024},  {HuZ`et`2024},  {Karniadakis`et`2021},  {Kim`L`2024},  {Raissi`et`2019},  {Ren`L`2024}} incorporate governing physical laws—such as Eq. (\ref{2.2}) -- directly into the training process by embedding the residuals of these equations into the loss function. This ensures that the neural network solutions remain consistent with the underlying physics.
		When PINNs are combined with neural operator frameworks  
\citep{{Faroughi`et`2024}, {Gosmani`et`2022}, {Wang`Y`2024}}, the resulting models can effectively learn complex material behavior, including nonlinear responses, microstructural heterogeneity, and nonlocal interactions, while maintaining strong generalization performance. Nevertheless, these approaches are typically limited to problems defined on finite computational domains and are not inherently suited for direct modeling of infinite media.

The procedure for generating surrogate models is outlined in Fig. 5. Block 1 (DNS, see $\bfcD^{\rm DNS}$,
Eq. (\ref{6.4})) was used
by  \citep{Fan`et`2023},  \citep{Jafarzadeh`et`2024},  \citep{You`et`2020},  \citep{You`et`2021}, and  \citep{You`et`2022}. In our CAMNN,
%approach, we instead use periodic or deterministic structure CMs (see Eqs. (\ref{3.7}) and
% (\ref{3.8})) and apply AMic or CMic tools to 
we instead 
obtain compressed datasets $\bfcD^{\rm p}$  (\ref{3.7}) or $\bfcD^{\rm d}$  (\ref{3.8}), respectively (Block 2). These serve as inputs to Block 3 
(Optimization), replacing the more data-intensive $\bfcD^{\rm DNS}$. This substitution significantly reduces dataset size and 
improves efficiency, requiring only minor adjustments between Blocks 2 and 3. Block 3 in Fig. 5 then yields either a single surrogate operator 
(e.g., $\bfK^*_b$ or $\bfcG$, Block 4, see  \citep{{Silling`2020},  {You`et`2020},  {You`et`2024}}) or a set 
of surrogate models (e.g., Eq. 
(\ref{6.5}) in Block 5). {\color{black} The only modification to the proposed approach is the replacement of Block 1 with Block 2 of the new compresed dataset; no adjustments to the existing Block 3 are required.}  

\vspace{-0.mm} \noindent \hspace{5mm} \parbox{11.2cm}{
\centering \epsfig{figure=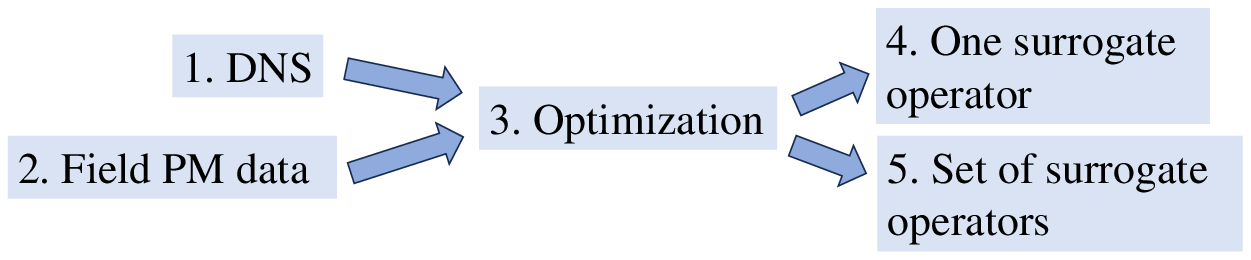, width=11.2cm}\\ \vspace{-119.mm}
\vspace{119.mm}
\vspace{-120.mm} \tenrm \baselineskip=8pt
{{\sc Fig. 5:} The scheme of obtaining of surrogate model set}}
\vspace{2.mm}

Periodic sructure CM with the PBC (\ref{2.22}) are totally dominated in analyses by the most popular ANN methods such as are NO  \citep{{Gosmani`et`2022},  {Hu`et`2024},  {Lanthaler`et`2024}}; PNO  \citep{{Jafarzadeh`et`2024}, {Jafarzadeh`et`2024b}}; PINN  \citep{{Eghbalpoor`S`2024},  {Faroughi`et`2024},  {Gosmani`et`2022},  {Haghighat`et`2021},  {Karniadakis`et`2021},  {Kingma`B`2014},  {Ning`et`2023},  {Paszke`et`2019},
 {Raissi`et`2019}}; and EINN, see   \citep{{Yu`Z`2024},{Yu`Z`2024b},{Zhou`Y`2024}}. 
A key advantage of the BFCS loading (\ref{2.4}) lies in its ability to generate datasets $\bfcD^{\rm p}$ and $\bfcD^{\rm d}$ while avoiding issues related to sample size, boundary layers, and edge effects. In addition, for periodic CMs, the CAM framework relies on solving problems (\ref{3.5}) and (\ref{3.6}) on the RVE, which may include several UCs, without requiring
 {\color{black} any specific PBCs (\ref{2.19}) at UC interfaces are lost. Consequently, the direct application of both asymptotic homogenization \citep{{Bakhvalov`P`1984}, {Fish`2014}} and computational homogenization methods \citep{{Geers`et`2010}, {Kanout`et`2009}, {Kouznetsova`et`2001}, {Matous`et`2017}, {Raju`et`2021}, {Terada`K`2001}}—which inherently assume PBCs (\ref{2.19})—becomes questionable.
Therefore, the revised RVE concept (see Definition 2.2), formulated under the general BFCS loading (\ref{2.5}), is fundamental to the CAMNN approach--and to its extension to the ML and NN techniques discussed in Section 6--when applied to both periodic and deterministic composite structures.}

{\color{black} 
Significant progress has been achieved in developing effective operators for both random 
\citep{{Drugan`2000},  {Drugan`2003},  {Drugan`W`1996}} and periodic
 \citep{{Ameen`et`2018},  {Kouznetsova`et`2004a},  {Kouznetsova`et`2004b},  {Smyshlyaev`C`2000}} 
structures
(see also \citep{Buryachenko`2022} for further references). 
Yet, all of these methods are confined to estimating predefined operators—most notably the ubiquitous fourth-order differential operator. If one dares to step outside this mold, for instance by replacing a strain-type model with a displacement-type strongly nonlocal model, the entire micromechanical problem must be rebuilt and solved from the ground up.  Moreover, selecting an alternative micromechanical method in this process poses an additional, nontrivial challenge.  The present approach breaks free from this constraint. Here, all micromechanical investigation culminates in the determination of a new dataset $\bfcD^{\rm p}$ (or $\bfcD^{\rm d}$). From that moment, the computational model becomes entirely agnostic to both the composite’s microstructure and the numerical method (FEA, FFT, or otherwise) used to obtain the dataset. This dataset $\bfcD^{\rm p}$ (or $\bfcD^{\rm d}$) can then be seamlessly approximated by ML\&NN techniques using any predetermined—or entirely ``a priori" undefined—surrogate model. No further micromechanical computation is ever required. In essence, once the dataset is built, the heavy machinery of micromechanics can be switched off—the future analysis is instant.}

\section{ Conclusion}
\vspace{.0mm}
\setcounter{equation}{0}
\renewcommand{\theequation}{7.\arabic{equation}}

{\color{black} 
To clarify the essence of the proposed approach, we briefly highlight the following key aspects: the novelty of the problem formulation, the presented solution, and the potential directions for its further development -- both theoretical and practical.
To the best of the author’s knowledge, the BFCS loading (\ref{2.5}) has not previously been applied in micromechanics -- whether for random or periodic structures -- likely due to an initial underestimation of its practical significance. However, the principal motivation for introducing BFCS loading lies in the fundamentally new opportunity it provides: namely, its role as a training parameter for the estimation of unspecified surrogate nonlocal operators (see Section  6).
This is achieved through several steps. First, the BFCS loading (\ref{2.5})  is employed to define a new RVE concept (see Definition 2.2), via a novel general translation-based averaging procedure (\ref{3.5}) and (\ref{3.6}).
%( \ref{3.5}) and (\ref{3.6}).
The required DNS for each specific grid $\bfLa_{\chi}$ in this averaging process can be carried out using any numerical method, particularly FFT-based techniques (see Section 5). The effective implementation of the new RVE concept relies critically on the accurate construction of the datasets $\bfcD^{\rm p}$ (\ref{3.7}) and $\bfcD^{\rm d}$ (\ref{3.8}). Only once these datasets have been precisely established can they be incorporated into ML\&NN frameworks (see Section 6).
Notably, the practical relevance of the ML-S equation (\ref{2.14}) under compact support loading conditions remains limited—unless enhanced through ML\&NN techniques. Without incorporating into ML\&NN components, the BFCS   (\ref{2.5})  equation offers minimal practical utility, with the exception of certain special cases, such as the laser heating problem discussed in \citep{Buryachenko`2025}. This explains why the ML-S equation (\ref{2.14}) —despite being conceptually simpler than the original L-S formulation (\ref{2.24})—has historically received little attention.
Only through a fully integrated framework -- combining BFCS 
loading (\ref{2.5}). 
 with the new RVE concept, refined datasets $\bfcD^{\rm r}$ or $\bfcD^{\rm d}$, and ML\&NN tools -- can the approach yield practical and predictive results. Symbolically, the proposed approach can be represented by the following sequence:
\BB
\label{7.1}
{\rm BFCF}\ \to \  {\rm RVE}\ \to \bfcD^p_{\chi k}\ \to \ \bfcD^p\ \to\  {\rm ML}\&{\rm NN}. 
\EE
This unified methodology (see \citep{Buryachenko`2025} for details) enables the prediction of a broad class of ``a priori" undefined surrogate operators—unlike the predefined effective nonlocal operators examined in \citep{{Ameen`et`2018},  %{Auriault`1983},  {Auriault`R`1993},
{Drugan`2000},  {Drugan`2003},  {Drugan`W`1996},{Kouznetsova`et`2004a},  {Kouznetsova`et`2004b},  {Smyshlyaev`C`2000}}. 
These operators encompass both macroscopic effective (nonlocal) properties and local concentration fields, while removing dependence on sample size, boundary conditions, and edge effects. These capabilities mark a substantial advancement and underscore the transformative potential of the ML-S equation when used in a data-driven context.
}

Regarding DNS on the specific grid of inclusions $\bfLa_{\chi}$, this paper extends FFT-based methods to the field of micromechanics for CMs with either periodic or deterministic structures, subjected to BFCS loading (\ref{2.5}). 
The advantages of the FFT approach in the field of computational homogenization are (i) its very efficient numerical
response, (ii) the reduced memory allocation needs, (iii) the possibility of using 2D/3D, and (iv) the periodicity of the fields, which does not require
the additional cost to impose periodicity in the FEM.
The FFT approach is proposed by the use of the most popular tools and concepts
exploited in the local elasticity of CMs (the modified 
Lippmann–Schwinger (\ref{2.14}) equation-based approaches) and adapted to the case of BFCS loading (\ref{2.5}).

Over the past three decades, following the seminal work of Moulinec and Suquet  \citep{Moulinec`S`1994}, numerous high-performance FFT-based algorithms have been developed to enhance convergence and accuracy in the analysis of microstructures with arbitrary phase contrast under finite deformations. These advancements have enabled the efficient simulation of nonlinear material behavior—such as plasticity, viscoplasticity, damage, fracture, and fatigue—on standard desktop computers, eliminating the need for high-performance computing clusters or supercomputers (see reviews in  \citep{Lucarini`et`2022},  \citep{Schneider`2021},  \citep{Segurado`et`2018}). 
These FFT solvers have also been extended to address multiphysics problems involving coupled mechanical, thermal, electrical, magnetic, and pyroelectric effects. For instance, they have been employed in chemo-thermo-mechanical modeling of batteries, including phase-field approaches to damage, where multiple fields interact through coupled partial differential equations.
A promising future direction would be to generalize these FFT methods—originally developed for PBC as in Eq.(\ref{2.19})—to accelerated solvers capable of handling BFCS-type loading conditions (Eq.(\ref{2.5})), such as those discussed in Section 5, with similarly improved convergence properties. 
It should be emphasized that the convergence behavior of the FFT schemes for the L–S equation (\ref{5.1}) and for the ML–S equation (\ref{5.3}) is identical, since the operator $\bfU^{(0)}$ is the same in both equations; they differ only in their free terms.

An emerging direction for improving predictive capabilities involves the use of ML and NN techniques to construct surrogate operators. However, despite their power, these methods often overlook fundamental micromechanical principles such as scale effects, boundary layers, and the concept of the RVE. To address this limitation, the proposed CAMNN approach generates fundamentally new, compressed datasets for both periodic and deterministic composite microstructures. 
The novelty of CAMNN lies in replacing the dataset $\bfcD^{\rm DNS}$ (\ref{6.4}) with either of the datasets $\bfcD^{\rm p}$ (\ref{3.7}) or $\bfcD^{\rm d}$ (\ref{3.8}), while the Block 3 Optimization in Fig. 5 remains unchanged.
The newly introduced RVE concept (Definition 2.2) represents a significant departure from classical definitions. It is revolutionary in that it does not depend on the constitutive behavior of the individual phases or the specific form of the surrogate operator being predicted. Instead, it is grounded in the behavior of field concentration factors within the composite phases. This abstraction makes the definition more flexible and broadly applicable across various modeling approaches.
Crucially, the generated datasets embed this generalized RVE concept as a core component, enabling seamless integration into existing ML and NN frameworks for the prediction of nonlocal surrogate operators. Incorporating this concept has the potential to significantly enhance both the reliability and generalizability of ML/NN-based models, particularly in complex systems where micromechanical accuracy is essential. By systematically eliminating size effects, boundary layers, and edge effects, the CAM approach ensures more robust and physically consistent predictions.

In summarizing, structuring, and generalizing the proposed approach, we highlight that it seamlessly integrates (three ingredients) a constitutive law agnostic RVE definition under BFCS loading, an FFT solver adapted to that loading, and ML/NN-based surrogates trained on compact, physics informed datasets. We begin with a novel RVE concept that’s defined through BFCS loading
(\ref{2.5}) -- this is entirely independent of any constitutive model, be it local, gradient-enhanced, peridynamic, or 
multiphysics (see classification of constitutive laws in  \citep{Maugin`2017}). This concept hinges on using field concentration factors within each phase (denoted in your notation as $\bfcD^p_{{\bf \chi}k}$ (\ref{3.7}) and $\bfcD^{d_j}_k$ 
(\ref{3.8}) to characterize the microstructure. Crucially, it doesn’t rely on a sample needing to be “large enough” or on specific phase behavior, making it universally applicable. Next, you drive this RVE into a numerical scheme based on the modified L-S equation (\ref{5.3}) and (\ref{5.4}) solved via FFT (the second ingredient). What makes our method stand out is replacing the usual homogeneous boundary conditions
(\ref{2.20}$_1$) or (\ref{2.20}$_2$) with BFCS loading (\ref{2.5}).
This transforms the standard FFT solver for L-S equation (\ref{5.1}) and (\ref{5.2}) —originally developed for linear, high-contrast matrices or polycrystals—into a more general tool that handles BFCS-driven, possibly nonlinear or nonlocal composites. Finally, rather than training machine learning models on full-field DNS data 
$\bfcD^{\rm DNS}$ (\ref{6.4}), we train on the reduced datasets generated via the new RVE and FFT steps. 
By feeding in $\bfcD^{\rm p}$ (\ref{3.7}) or $\bfcD^{\rm d}$ (\ref{3.8}) -- 
as opposed to massive DNS outputs -- our CAMNN method builds surrogate operators with much smaller, noise-free inputs, while the optimization framework remains the same (the third ingredient, see Block 3 Optimization in Fig. 5). Altogether, the innovation lies in how the BFCS-driven RVE + FFT yields compact, constitutive-independent microstructure descriptors that feed directly into ML/NN models (a set of surrogate operators, see Section 6). This creates a modular, robust pipeline that generalizes across local/nonlocal, linear/nonlinear, and multi-physics CMs.

Each of the three principal components of the proposed methodology addresses a remarkably broad spectrum of problems. These problems, despite their complexity or specificity, can often be effectively treated through relatively minor modifications or strategic reformulations of well-established problem classes. In the present work, the author has deliberately focused on providing a conceptual and schematic presentation of these reductions, emphasizing the fundamental ideas that underlie the generalization of classical approaches. Readers who are interested in the technical nuances, computational strategies, and practical implementations of the proposed framework are encouraged to pursue these aspects independently. A comprehensive treatment of such applications—while undoubtedly important and promising—falls outside the scope of the current theoretical investigation and will be addressed in future work or in applied extensions of the present study.

\noindent{\bf Acknowledgments:}

%The author acknowledges Dr. Stewart A. Silling for the helpful
%comments, fruitful personal discussions, and encouragement suggestions.
%The author  acknowledges the reviewers for the
%encouraging comments that initiated a significant correction of the manuscript.
%The author acknowledges the reviewers for the
%encouraging comments that initiated a significant correction of the manuscript.
Permission to reproduce Figs. 4a and 4b from \cite{Silling`et`2024} was granted by Springer Nature (License Number 6044880317237).

%{\color{black}The author also acknowledges
%the reviewers for the encouraging comments that initiated improvement
% of the manuscript.}

%{\color {black} Both the helpful comments of reviewers and their encouraging recommendations are gratefully acknowledged.
%{\color{black} Both the helpful comments of reviewers and their encouraging
%recommendations are g`ratefully acknowledged.}

%\citep{Ameenet`2018}  aa bb
%\citep{{Ameenet`2018},  {Amidror`2013}}
%\citep{Amidror`2013}

%\begin{thebibliography}{99}


\begin{thebibliography}{99}

{\baselineskip=3pt
\parskip=0.1pt \tenrm
\tenrm
{

\bibitem{Fish`2014}
Fish J {\tenit Practical Multiscaling}. Chichester: John Wiley \& Sons, 2014.

\bibitem{Ghosh`2011}
Ghosh, S. {\tenit Micromechanical Analysis and Multi-Scale Modeling Using the Voronoi Cell Finite
Element Method (Computational Mechanics and Applied Analysis)}. Boca Raton: CRC Press, 2011.

\bibitem{Zohdi`W`2008}
Zohdi TI, Wriggers P. 
{\tenit Introduction to Computational Micromechanics.}
Berlin: Springer, 2008.

\bibitem{Bakhvalov`P`1984}
Bakhvalov NS, Panasenko GP {\tenit Homogenisation: Averaging Processes in Periodic Media}. Nauka, Moscow (in Russian; English translation: Kluwer, 1989), 1984.

\bibitem{Buryachenko`2024}
Buryachenko V. A. Peridynamic micromechanics of composites: a review. 
{\tenit J. Peridynamics and Nonlocal Modeling}, 2024; {\tenbf 6}: 531–601 

\bibitem{Kouznetsova`et`2001}
Kouznetsova VG, Brekelmans WAM, Baaijens FPT 
An approach to micro-macro modeling of heterogeneous materials, {\tenit Comput. Mech.}  2001; {\tenbf 27}: 37-–48

\bibitem{Matous`et`2017}
Matouš K, Geers MGD, Kouznetsova VG, Gillman A 2017; A review of predictive
nonlinear theories for multiscale modeling of heterogeneous materials.
{ \tenit J. Comput. Physics}, 2017; {\tenbf 330}: 192–220

\bibitem{Terada`K`2001}
Terada K, Kikuchi N.A class of general algorithms for multi-scale analyses of heterogeneous media, {\tenit Comput. Methods Appl. Mech. Eng.} 2001;  { \tenbf 190}: 5247–-5464

\bibitem{Geers`et`2010}
Geers, MGD, Kouznetsova, VG, Brekelmans, WAM. Multi-scale computational
homogenization: Trends and challenges. {\tenit J. Comput. Applied Mathematics}. 2010;
{\tenbf 234}: 2175-2182.

\bibitem{Kanout`et`2009}
Kanout\'e P, Boso DP, Chaboche LJ, Schrefler BA. 
Multiscale methods for composites: a review
{\tenit Archives of Computational Methods in Engineering}, 2009; {\tenbf 16}: 31–75.

\bibitem{Raju`et`2021}
Raju K, Tay TE, Tan VBC.
A review of the FE2 method for composites
{\tenit Multiscale and Multidisc. Modeling Experiments and Design}, 2021;
{\tenbf 4}: 1–24

\bibitem{Moulinec`S`1994}
Moulinec H, Suquet P. Fast numerical method for computing the linear and nonlinear properties
of composites. {\tenit C. R. Acad. Sci., Paris}, (1994;  {\tenbf 318}: 1417–23

\bibitem{Moulinec`S`1998}
Moulinec H, Suquet P.  A numerical method for computing the overall response of nonlinear
composites with complex microstructure. {\tenit Comput. Methods Appl. Mech. Eng.}, 1998;  {\tenbf 157}: 69–94

\bibitem{Michel`et`2001}
Michel J C, Moulinec H, Suquet P. A computational scheme for linear and non-linear composites
with arbitrary phase contrast. {\tenit Int. J. Numer. Methods Eng.} 2001;  {\tenbf 52}: 139–60

\bibitem{Brisard`D`2010}
Brisard S., Dormieux L. FFT-based methods for the mechanics of composites: a general
variational framework. {\tenit Comput. Mater. Sci.} 2010;  {\tenbf 49}: 663–71

\bibitem{Zeman`et`2010}
Zeman J, Vond$\breve{\rm r}$ejc J, Nov\'ak J, Marek I. Accelerating an FFT-based solver for numerical
homogenization of periodic media by conjugate gradients. {\tenit J. Comput. Phys.} 2010;  {\tenbf 229}: 8065–71

\bibitem{Wicht`et`2021}
Wicht D, Schneider M., B\"ohlke T. Anderson-accelerated polarization schemes for FFT-based
computational homogenization. {\tenit Int. J. Numer. Methods Eng.}, 2021; {\tenbf 122}: 2287–311

\bibitem{Wang`et`2024}
Wang B, Li M, Fang G, Hu J, Ye J, Meng S. 
A novel FFT framework with coupled non-local elastic-plastic damage
model for the thermomechanical failure analysis of UD-CF/
PEEK composites. {\tenit Compos. Science Technol.}, 2024; {\tenbf 251}: 110540

\bibitem{LiM`et`2024}
Li M, Wang B, Hu J, Li G, Ding P, Ji C. 
Artificial neural network-based homogenization model for predicting multiscale thermo-mechanical properties of woven composites
{\tenit Int. J. Solids Struct.}, 2024; {\tenbf 301}: 112965

\bibitem{deGeus`et`2017}
de Geus T W J, Vond$\breve{\rm r}$ejc J, Zeman J, Peerlings RHJ, Geers MGD. Finite strain FFT-based
non-linear solvers made simple. {\tenit Comput. Methods Appl. Mech. Eng.} 2017; {\tenbf 318}, 412–30

\bibitem{Zeman`et`2017}
Zeman J, de Geus TWJ, Vondrejc J, Peerlings RHJ, Geers MGD. A finite element perspective
on nonlinear FFT-based micromechanical simulations. {\tenit Int. J. Numer. Methods Eng.} 2017; {\tenbf 111}: 903–26

\bibitem{Lucarini`S`2019}
Lucarini S, Segurado J. DBFFT: a displacement-based FFT approach for non-linear homogenization
of the mechanical behavior. {\tenit Int. J. Eng. Sci.}, 2019; {\tenbf 144}, 103-131

\bibitem{Schneider`2021}
Schneider M. A review of nonlinear FFT-based computational
homogenization methods. {\tenit Acta Mech}, 2021; {\tenbf 232}, 2051–2100

\bibitem{Segurado`et`2018}
Segurado J, Lebensohn R A, Llorca J. Computational homogenization of polycrystals
{\tenit Advances in Applied Mechanics} 2018; {\tenbf 51}: 1–114

\bibitem{Lucarini`et`2022}
Lucarini S, Upadhyay MV, Segurado J. FFT-based approaches in micromechanics:
fundamentals, methods, and applications. {\tenit Modelling Simul. Mater. Sci. Eng}. 2022; {\tenbf 30}: 023002 (97pp.)

\bibitem{Hill`1963}
Hill R. Elastic properties of reinforced solids: some theoretical principles. {tenit J Mech
Phys Solids},  1963; 11:357–372

\bibitem{Bargmann`et`2018} %%5
Bargmann~S, Klusemann~B, Markmann~J, Schnabel~JE, Schneider~K, Soyarslan~C, Wilmers~J
Generation of 3D representative volume elements for heterogeneous materials: A review. 
{\it Progress in Materials Science}, 2018; { 96}: 322--384

\bibitem{Kanit`et`2003} %!!XX
Kanit~T, Forest~S, Galliet~I, Mounoury~V, Jeulin~D
Determination of the size of the representative volume
element for random composites: statistical
and numerical approach. {\it Int J Solids Struct,} 2003; { 40}: 3647--3679

\bibitem{Moumen`et`2021} %%^6
Moumen~AE, Kanit~T, Imad~A 
Numerical evaluation of the representative volume element for
random composites. {\it European Journal of Mechanics / A Solids}, 2021; { 86}: 104181

\bibitem{Ostoja`et`2016}
Ostoja-Starzewski M, Kale S, Karimi P, Malyarenko A, Raghavan B, Ranganathan SI,
Zhang J Scaling to RVE in random media. {tenit Adv. Appl. Mech.}, 2016; { 49}: 111–211


\bibitem{Drugan`2000} %!!XIII 
Drugan~WJ Micromechanics-based variational estimations 
for a higher-order nonlocal constitutive equation and optimal choice 
of effective moduli for elastic composites.
{\tenit J Mech Phys Solids,} 2000; { 48}: 1359--1387

\bibitem{Drugan`2003} %%%
Drugan~WJ Two exact micromechanics-based nonlocal constitutive equations for random linear elastic composite materials.
{\tenit J Mech Phys Solids}, 2003; { 51}: 1745--1772 

\bibitem{Drugan`W`1996} %!!VII XII XIII 
Drugan~WJ, Willis~JR A micromechanics-based
nonlocal constitutive equation and estimates of representative volume elements
for elastic composites. {\tenit J Mech Phys Solids,} 1996; { 44}: 497--524

\bibitem{Ameen`et`2018}
Ameen MM, Peerlings RHJ, Geers MGD. A 
quantitative assessment of the scale separation limits of classical and
higher-order asymptotic homogenization
{\tenit European J. Mech. A. Solids},  2018; {\tenbf 71}: 89–100

\bibitem{Auriault`1983}
Auriault J Effective macroscopic description for heat conduction in periodic composites. 
{\tenit Int. J. Heat Mass Tranfer}, 1983; {\tenbf 26}: 861--869.

\bibitem{Kouznetsova`et`2004a}
Kouznetsova V, Geers M, Brekelmans W. 
Size of a representative volume element in a second-order computational homogenization framework.
{\tenit Int. J. Multiscale Comput. Eng.}, 2004; { 2}: 575--598

\bibitem{Kouznetsova`et`2004b}
Kouznetsova V, Geers M, Brekelmans W. 
Multi-scale second-order computational homogenization of multi-phase materials: a nested finite element solution strategy.
{\tenit Comput. Methods Appl. Mech. Eng.}, 2004; { 193}: 5525--5550

\bibitem{Smyshlyaev`C`2000} %!!VII XIII 
Smyshlyaev VP, Cherednichenko KD A rigorous derivation of 
strain gradient effects in the overall behavior of periodic heterogeneous
media. {\it J Mech Phys Solids,} 2000; {\bf 48}:1325--1357

\bibitem{Silling`2020}
Silling S. Propagation of a stress pulse in a heterogeneous elastic bar. {\tenit Sandia Report
SAND2020-8197}, Sandia National Laboratories, 2020.


\bibitem{You`et`2020}
You H, Yu Y, Silling S, D’Elia M Data-driven learning of nonlocal models: from
high-fidelity simulations to constitutive laws. { \tenit arXiv:2012.04157}, 2020

\bibitem{You`et`2024}
You H, Yu Y, Silling S, D'Eliac M. 
Nonlocal operator learning for homogenized models:
from high-fidelity simulations to constitutive laws.
{\tenit J. Peridynamics Nonlocal Modeling},
https://doi.org/10.1007/s42102-024-00119-x, 2024.

\bibitem{Li`et`2003}
Li Z, Kovachki N, Azizzadenesheli K, Liu B, Bhattacharya K, Stuart A, Anandkumar, A. Neural
operator: Graph kernel network for partial differential equations, arXiv preprint arXiv:2003.03485, 2003.

\bibitem{Lanthaler`et`2024}
Lanthaler S, Li Z, Stuart AM. Nonlocal and nonlinearity imply universality in operator learning.
arXiv:2304.13221v2, 2024.

\bibitem{Gosmani`et`2022}
Goswami S, Bora A, Yu Y, Karniadakis GE. Physics-Informed
Neural Operators, arXiv preprint arXiv:2207.05748, 2022. 


\bibitem{Hu`et`2024}
Hu H, Qi L, Chao X. 2024)
Physics-informed Neural Networks (PINN; for computational solid mechanics: Numerical frameworks and applications
Author links open overlay panel. {tenit Thin-Walled Structures}, 2024;  112495

\bibitem{Kumara`Y`2023}
Kumara H, Yadav N. 
Deep learning algorithms for solving differential equations: a
survey. {\tenit J. Experimental \& Theoret. Artificial Intelligence}: 
https://doi.org/10.1080/0952813X.2023.212356, 2023.

{\color{black} \bibitem{Jafarzadeh`et`2024}
Jafarzadeh S, Silling S, Liu N, Zhang Z, Yu Y. 
Peridynamic neural operators: a data-driven nonlocal constitutive model for complex material responses.
{\tenit arXiv preprint arXiv:2401.06070}}, 2024.

\bibitem{Jafarzadeh`et`2024b}
Jafarzadeh S, Silling S, Zhang L, Ross C, Lee CH, Rahman SM, Wang S, Yu Y. 
Heterogeneous peridynamic neural operators: discover biotissue constitutive law and microstructure from digital image correlation measurements. {\tenit ArXiv preprint arXiv:2403.18597}, 2024.

\bibitem{Raissi`et`2019}
Raissi M, Perdikaris P, Karniadakis GE. 
Physics-informed neural networks: A deep learning framework for solving forward and
inverse problems involving nonlinear partial differential equations. {\tenit J. Comput. Phys.}, 2019; { 378}: 686–707

\bibitem{Karniadakis`et`2021}
Karniadakis GE, Kevrekidis IG, Lu L, Perdikaris P, Wang S, Yang L. 
Physics- informed machine learning.
{\tenit Nature Reviews Physics}, https://doi.org/10.1038/
s42254-021-00314-5, 2021;

\bibitem{Faroughi`et`2024}
Faroughi SA, Pawar NM, Fernandes C, Raissi M, Das S, Kalantari NK, Kourosh Mahjour S. 
Physics-guided, physics-informed, and physics-encoded neural networks and operators in scientific computing: Fluid and solid mechanics
{J. Computing and Information Science}, 2024; { 24}: 040802

\bibitem{Wang`Y`2024}
Wang X, Yin Z-Y. 
Interpretable physics-encoded finite element network to handle
concentration features and multi-material heterogeneity
in hyperelasticity.
{\tenit Comput. Meth. Applied Mech. Engng}, 2024; {tenit 431}: 117268

\bibitem{Mura`1987}
Mura T.  {\tenit Micromechanics of Defects in Solids (Mechanics of Elastic and Inelastic Solids)} 2nd edn
Berlin: Springer, 1987.

\bibitem{Milton`2002}
Milton GW.
{\tenit The theory of Composites}. Cambridge University Press, Cambridge, UK, 
2022.

\bibitem{Michel`et`1999}
Michel J, Moulinec H, Suquet P.
Effective properties of composite materials with periodic
microstructure: a computational approach. {\tenit Comput. Methods Appl. Mech. Engrg.}, 1999: {\tenbf 172}, 109--143

\bibitem{Buryachenko`2022}
Buryachenko VA. {\tenit Local and Nonlocal Micromechanics of Heterogeneous Materials.} Springer, NY 2022.

\bibitem{Konig`et`1991} %!!V 
K\"onig D, Carvajal-Gonzalz S, Downs AM, Vassy J, Rigaut JP.
Modelling and analysis of
3-D arrangements of particles by point process with examples of application to biological data obtained by
confocal scanning light microscopy. 
{\it J Microscopy,}  1991; {\bf 161}:405--433


\bibitem{Ohser`M`2000} %!!V 
Ohser J, F. M\"ucklich F. {\it Statistical Analysis of Microstructures in Material Science.} 
John Wiley \& Sons, Chichester, 2000.

\bibitem{Torquato`2002} %!!V 
Torquato S. Statistical description of microstructures. {\it Annu Rev Mater Res,} 2002;  {\bf 32}:77--111}}

\bibitem{Buryachenko`2024b}
Buryachenko V. A. Critical analyses of RVE concepts in local and peridynamic micromechanics of composites. {\tenit J. Period. Nonlocal Modeling} (submitted), https://arxiv.org/abs/2402.13908v5 (71pp, 336refs.), 2024.

\bibitem{Buryachenko`2007}
Buryachenko VA. {\tenit Micromechanics of Heterogeneous Materials}. Springer, NY, 2007.

\bibitem{Silling`et`2024}
Silling SA, Jafarzadeh S, Yu Y. 
Peridynamic models for random media found by coarse graining
{\tenit J. Peridynamics and Nonlocal Modeling}, 2024; {\tenbf 6}

\bibitem{Buryachenko`2023}
Buryachenko VA.
Effective nonlocal behavior of peridynamic random structure
composites subjected to body forces with compact support and related prospective problems.
{\tenit Math. Mech. of Solids}, 2023; {\tenbf 28}: 1401-1436

\bibitem{Buryachenko`2023a}
Buryachenko V Effective displacements of peridynamic heterogeneous bar loaded by body force with compact support.
{\tenit J. Multiscale Comput. Enging}, 2023b;  {\tenbf 21}: 27--42

\bibitem{Amidror`2013}
Amidror I. {\tenit Mastering the Discrete Fourier Transform in One, Two or Several Dimensions.} Springer-Verlag London, 2013.

\bibitem{Marks`2009}
Marks II R. J. {\tenit Handbook of Fourier Analysis and its Applications}. Oxford University
Press, NY, 2009.

\bibitem{Briggs`H`1995}
Briggs, WL, Henson, VE. {\tenit The DFT: An Owner’s Manual for the Discrete Fourier
Transform}. SIAM, Philadelphia, 1995.

\bibitem{Brigham`1988}
Brigham, EO. {\tenit The Fast Fourier Transform and Its Applications}. Prentice-Hall, NJ, 1988.

\bibitem{Cooley`T`65}
Cooley JW, Tukey JW. An algorithm for the machine calculation of complex Fourier series.
{\tenit Math. Comput.}, 1965; {\tenbf 19}: 297–301

\bibitem{Buryachenko`2023j}
Buryachenko V. A. Fast Fourier transform in peridynamic micromechanics of composites {\tenit Math. Mech. of Solids}, 2024; {\tenbf 29}, https://doi.org/10.1177/10812865241236878

\bibitem{Barrett`et`1994}
Barrett R, Berry M, Chan TF, Demmel J, Donato J, Dongarra J, Eijkhout V., Pozo R, Romine C, der
Vorst HV. {\tenit Templates for the Solution of Linear Systems: Building Blocks for Iterative Methods}, 2nd Edition, SIAM, 1994.

\bibitem{Vondrejc`et`2012}
Vond$\breve{\rm r}$ejc J, Zeman J, Marek I. Analysis of a fast Fourier transform-based method for modeling
of heterogeneous materials. {\tenit Large-Scale Scientific Computing}. Eds.I Lirkov, S Margenov, J
Wa\'sniewski (Berlin: Springer), 515—522, 2012.

\bibitem{Lahellec`et`2003}
Lahellec N, Michel J C, Moulinec H., Suquet P. Analysis of inhomogeneous materials at large
strains using fast Fourier transforms. {\tenit IUTAM Symposium on Computational Mechanics of Solid
Materials at Large Strains} (Berlin: Springer; pp. 247–58, 2003.

\bibitem{Bellens`et`2024}
Bellens S, Guerrero P, Vandewalle P, Dewulf W. 
Machine learning in industrial X-ray computed tomography – a review
{\tenit CIRP J. Manuf. Science Techn.}, 2024; {\tenbf 51}: 324–341

\bibitem{Echlin`et`2014} %%VI
Echlin~MP, Lenthe~WC, Pollock~TM.
Three-dimensional sampling
of material structure for property modeling and design. {\tenit Integ.Mater.
Manuf. Innov.}, 2014; {\tenbf 3}:1--14

\bibitem{Balay`et`2016} Balay, S., Abhyankar, S., Adams, M., Brown, J., Brune, P. {\tenit et al.} 
{\tenit PETSc users manual 3.7 Technical Report}, Argonne National Lab. (ANL), Argonne, IL, 2016.

\bibitem{Plimpton`et`2018}
Plimpton, S., Kohlmeyer, A., Coffman, P., Blood, P. 
{\tenit fftMPI, a library for performing 2d and 3d FFTs in parallel.} Computer software. Sandia National Lab. (SNL-NM; https://www.osti.gov//servlets/purl/1457552. USDOE. 25 Apr. 2018. Web. doi:10.11578/dc.20201001.68, 2018.

\bibitem{Eyre`M`1999}
Eyre, D.J, Milton, G.W. A fast numerical scheme for computing the response of composites using
grid refinement. {\tenit Eur. Phys. J.: Appl. Phys.}, (1999;  {\tenbf 6}: 41–47

\bibitem{Kingma`B`2014}
Kingma DP, Ba J. Adam: A method for stochastic optimization. {\tenit arXiv:1412.6980}, 2014.

\bibitem{Fan`et`2023}
{\color{black} Fan Y, D’Elia M, Yu Y, Najm HN, Silling S. Bayesian nonlocal operator regression: A data-driven
learning framework of nonlocal models with uncertainty quantification. {\tenit J. Engig
Mech.}, 2023; {\tenbf 149}: 04023049.}

\bibitem{You`et`2021}
You H, Yu Y, Trask N, Gulian M, D’Elia M. Data-driven learning of robust nonlocal
physics from high-fidelity synthetic data. {\tenit Computer Methods Applied Mech. Engineering}, 2021; {\tenbf 374}: 113553

\bibitem{You`et`2022}
You H, Zhang Q, Ross CJ, Lee CH, Yu Y. 
Learning deep implicit Fourier neural operators (IFNOs; with applications to heterogeneous material modeling. 2022;
{\tenit Com.r Meth. Appl. Mech. Engng.}, {\tenbf 398}: 115296

\bibitem{Silling`et`2003}
Silling SA, Zimmermann M, Abeyaratne R. Deformation of a peridynamic bar.
{\tenit J. Elasticity}, 2003; { \tenbf 73}: 173--190.

\bibitem{HuZ`et`2024}
Hu Z, Daryakenari NA, Shen Q, Kawaguchi K.
Karniadakis GE. 
State-space models are accurate and efficient neural operators for
dynamical systems. arXiv:2409.03231, 2024.

\bibitem{Cuomo`et`2022}
Cuomo S, Schiano V, Cola D, Giampaolo F, Rozza G, Raissi M, Piccialli F. Machine learning through physics–informed
neural networks: where we are and what’s next. 
{\tenit J. Scientific Computing},2022; {\tenbf 92}:88 (62pp)

\bibitem{Haghighata`et`2021}
Haghighata E, Raissi M, Moure A, Gomez H, Juanes R. 
A physics-informed deep learning framework for inversion and
surrogate modeling in solid mechanics
{\tenit Comput. Methods Appl. Mech. Engrg.}, 2021; {\tenbf 379}: 113741

\bibitem{Harandi`et`2024}
Harandi A, Moeineddin A, Kaliske M, Reese S. 
Mixed formulation of physics‐informed neural networks for thermo‐mechanically coupled systems and heterogeneous domains.
{\tenit Int J Numer Methods Eng}, 2024; {\tenbf 125}: e7388

\bibitem{Kim`L`2024}
Kim D, Lee J. 
A review of physics informed neural networks for multiscale analysis
and inverse problems.
{\tenit Multiscale Science and Engineering}, 2024;  {\tenbf 6}: 1–11

\bibitem{Ren`L`2024}
Ren X, Lyu X. 
Mixed form-based physics-informed neural networks for performance
evaluation of two-phase random materials
{\tenit Engng Appl. Artificial Intelligence}, 2024; {\tenbf 127}: 107250

\bibitem{Eghbalpoor`S`2024}
Eghbalpoor R, Sheidaei A. 
A peridynamic-informed deep learning model for brittle damage prediction
{\tenit Theoret. Appl. Fracture Mech.}, 2024; {\tenbf 131}: 104457

\bibitem{Haghighat`et`2021}
Haghighat E, Bekar AC, Madenci E, Juanes R. A nonlocal physics-informed deep learning framework using the peridynamic
differential operator. {\tenit Comput. Methods Appl. Mech. Engrg.} 2021;  {\tenbf 385}: 114012

\bibitem{Ning`et`2023}
Ning L, Cai Z, Dong H, Liu Y, Wang W. 
A peridynamic-informed neural network for continuum elastic
displacement characterization
{\tenit Comput. Methods Appl. Mech. Engrg.}, 2023; {\tenbf 407}: 115909

\bibitem{Paszke`et`2019}
Paszke A, Gross S, Massa F et al. PyTorch: an imperative
style, high-performance deep learning library. {\tenit Adv Neural Inf Process
Syst}, 2019; {\tenbf 32}:8024--8035

\bibitem{Yu`Z`2024}
Yu X-L, Zhou X-P. 
A nonlocal energy-informed neural network for peridynamic
correspondence material models. {\tenit Engng Anal. Boundary Elements}, 2024; {\tenbf 160}, 273--297

\bibitem{Yu`Z`2024b}
Yu X-L, Zhou X-P. 
A nonlocal energy-informed neural network based on peridynamics
for elastic solids with discontinuities {\tenit Computational Mechanics},  2024; {\tenbf 73}: 233--255

\bibitem{Zhou`Y`2024}
Zhou X-P, Yu X-L. 
Transfer learning enhanced nonlocal energy-informed neural
network for quasi-static fracture in rock-like materials.
{\tenit Comput. Methods Appl. Mech. Engrg.} https://doi.org/10.1016/j.cma.2024.117226, 2024.

\bibitem{Buryachenko`2025}
Buryachenko V. A. 2025; Unified Micromechanics Theory of Composites 
{\tenit https:// arxiv.org/abs/2503.14529} (89pp, 514refs.)

\bibitem[Maugin(2017)]{Maugin`2017}
Maugin GA. {\tenit Non-Classical Continuum Mechanics. A Dictionary.} Springer Nature,
Singapore, 2017


\end{thebibliography}
\end{document}